\documentclass[fleqn,usenatbib]{mnras}

\usepackage{newtxtext,newtxmath}
\usepackage[T1]{fontenc}

\DeclareRobustCommand{\VAN}[3]{#2}
\let\VANthebibliography\thebibliography
\def\thebibliography{\DeclareRobustCommand{\VAN}[3]{##3}\VANthebibliography}

\usepackage{graphicx}	% Including figure files
\usepackage{amsmath}	% Advanced maths commands
\usepackage{xcolor}
\usepackage{textgreek}

\graphicspath{Immagini/}

\newcommand{\solarM}{\,\mathrm{M}_\odot}

\newcommand{\E}[1]{\times10^{#1}}
\newcommand{\nH}{\,\mathrm{cm}^{-2}}
\newcommand{\ergcms}{\,\mathrm{erg}\,\mathrm{cm}^{-2}\,\mathrm{s}^{-1}}
\newcommand{\asec}{\,\mathrm{arcsec}}

\newcommand{\magnitude}{\,\mathrm{mag}}
\newcommand{\ergs}{\,\mathrm{erg}\,\mathrm{s}^{-1}}

\newcommand{\LX}{L_\mathrm{X}}
\newcommand{\FX}{F_\mathrm{X}}
\newcommand{\xmm}{\emph{XMM-Newton}}
\newcommand{\swift}{\emph{Swift}}
\newcommand{\chandra}{\emph{Chandra}}
\newcommand{\erosita}{\emph{eROSITA}}
\newcommand{\nustar}{\emph{NuSTAR}}
\newcommand{\belz}{J0456}
\newcommand{\salt}{\emph{SALT}}
\newcommand{\rxte}{\emph{RXTE}}
\newcommand{\belzlong}{4XMM\,J045626.3-694723}
\newcommand{\obsid}{0841660501}

\title[A new candidate magnetar in LMC]{Discovery of a magnetar candidate X-ray pulsar in the Large Magellanic Cloud\thanks{Based on observations obtained with XMM-Newton, an ESA science mission
with instruments and contributions directly funded by
ESA Member States and NASA.}}

\author[M. Imbrogno et al.]{
M. Imbrogno,$^{1,2}$\thanks{E-mail: matteo.imbrogno@inaf.it}
G.~L. Israel,$^{2}$
G.~A. Rodr\'{i}guez Castillo,$^{3}$
D.~A.~H. Buckley,$^{4,5}$
F. Coti Zelati,$^{6,7}$
\newauthor
N. Rea,$^{6,7}$
I.~M. Monageng,$^{4,5}$
P. Casella,$^{2}$
L. Stella,$^{2}$
F. Haberl,$^{8}$
P. Esposito,$^{9,10}$
F. Tombesi,$^{1,2,11,12,13}$
\newauthor
A. De Luca,$^{10}$
and A. Tiengo$^{9,10}$
\\
$^{1}$Dipartimento di Fisica, Università degli Studi di Roma “Tor Vergata”, via della Ricerca Scientifica 1, I-00133 Rome, Italy\\
$^{2}$INAF--Osservatorio Astronomico di Roma, via Frascati 33, I-00078 Monteporzio Catone, Italy\\
$^{3}$INAF/IASF Palermo, via Ugo La Malfa 153, I-90146 Palermo, Italy\\
$^{4}$South African Astronomical Observatory, PO Box 9, Observatory, Cape Town 7935, South Africa\\
$^{5}$Department of Astronomy, University of Cape Town, Private Bag X3, 7701 Rondebosch, South Africa\\
$^{6}$Institute of Space Sciences (ICE, CSIC), Campus UAB, Carrer de Can Magrans s/n, E-08193 Barcelona, Spain\\
$^{7}$Institut d’Estudis Espacials de Catalunya (IEEC), Carrer Gran Capità 2–4, E-08034 Barcelona, Spain\\
$^{8}$Max-Planck-Institut f\"{u}r extraterrestrische Physik, Gie\ss{}enbachstra\ss{}e 1, 85748 Garching, Germany\\
$^{9}$Scuola Universitaria Superiore IUSS Pavia, Palazzo del Broletto, piazza della Vittoria 15, I-27100 Pavia, Italy\\
$^{10}$INAF/IASF Milano, via A. Corti 12, I-20133 Milan, Italy\\
$^{11}$Department of Astronomy, University of Maryland, College Park, MD 20742, USA\\
$^{12}$NASA Goddard Space Flight Center, Greenbelt, MD 20771, USA\\
$^{13}$INFN – Roma Tor Vergata, via della Ricerca Scientifica 1, I-00133 Rome, Italy\\
}

\date{Accepted XXX. Received YYY; in original form ZZZ}

\pubyear{2023}

\begin{document}
\label{firstpage}
\pagerange{\pageref{firstpage}--\pageref{lastpage}}
\maketitle

% Abstract of the paper
\begin{abstract}

During a systematic search for new X-ray pulsators in the \xmm\ archive, we discovered a high amplitude ($PF\simeq86\%$) periodic 
%LS modulation at a period of 
($P\simeq7.25\,\mathrm{s}$) modulation 
in the X-ray flux of \belzlong\ (\belz\ hereafter), a previously unclassified source in the Large Magellanic Cloud (LMC). 
%LS time scale of the pulsations 
The period of the modulation is strongly suggestive 
%LS that \belz\ is 
of a spinning neutron star (NS). 
%LS The new pulsar is highly variable and detected only 
%LS once (out of 6 observations) 
The source was detected only during one
out of six observations in 2018-2022. Based on an absorbed power-law spectral model with photon slope of $\Gamma\simeq 1.9$, 
%it has been detected only once in 25 years, despite the fact the field has been observed by multiple X-ray observatories (sei sicuro di questa affermazione ? io ricordavo che era una posizione poco coperta dalle osservazioni). 
we derive a  0.3--10\,keV luminosity of $\LX\simeq2.7\E{34}\ergs$ for
%LS Assuming the 
a distance of 50\,kpc. 
The X-ray properties of \belz\ are at variance 
%LS) and an absorbed power-law spectrum with $\Gamma\simeq1.93$, we derive a 0.3--10\,keV luminosity of $\LX\simeq2.7\E{34}\ergs$. 
with those of 
%LS Although the vast majority of 
variable LMC X-ray pulsars hosted in high-mass X-ray binary systems with a Be-star companion.
%the luminosity and high pulsed fraction ($\simeq86\%$) of \belz\ are at odds with the values shown by these systems in both Magellanic Clouds (MCs). 
Based on \salt\ spectroscopic observations of the only optical object that matches the X-ray uncertainty region, we cannot completely 
%exclude 
rule out 
%ls the possibility, though unlikely, 
that \belz\ is a NS accreting from a late-type (G8-K3) star, an as-yet-unobserved  binary evolutionary outcome 
%LS that has not yet been observed 
in the MCs. % at these spin periods. 
We show that the source properties are in better agreement with 
%LSthe source being a 
those of magnetars. \belz\ may thus be second known magnetar in the LMC after SGR\,0526–66.
%ls quests frase la salterei, e' piu' da discussione che da abstract Future measurements of secular spin period changes may corroborate this interpretation. 
%If confirmed, it would represent the second known magnetar in the LMC after SGR\,0526–66. %, one of the three objects for which a Giant Flare has been detected
%Based on SALT spectroscopic observations of the only optical object that matches the X-ray uncertainty region, we cannot completely exclude the possibility, though unlikely, that \belz\ is instead a NS accreting from a late-type (G8-K3) star, a binary evolutionary outcome that was never observed before in the MCs.

%However, a system like this has never been observed in the MCs and it is highly improbable given the recent star formation episode.   

%This is a simple template for authors to write new MNRAS papers.
%The abstract should briefly describe the aims, methods, and main results of the paper.
%It should be a single paragraph not more than 250 words (200 words for Letters).
%No references should appear in the abstract. 
\end{abstract}

% Select between one and six entries from the list of approved keywords.
% Don't make up new ones.
\begin{keywords}
stars: neutron -- galaxies: individuals: Large Magellanic Cloud
\end{keywords}

%%%%%%%%%%%%%%%%%%%%%%%%%%%%%%%%%%%%%%%%%%%%%%%%%%

%%%%%%%%%%%%%%%%% BODY OF PAPER %%%%%%%%%%%%%%%%%%

\section{Introduction}\label{sec:intro}

The Small and Large Magellanic Clouds (SMC and LMC, respectively) are the two best-studied star-forming satellite galaxies of the Milky Way. Thanks to their recent star formation bursts (in the inner regions)  occurred $\sim$6--25\,Myr ago in the LMC and $\sim$25--60\,Myr ago in the SMC \citep{Antoniou:2010,Antoniou:2016}, the MCs host a high number of X-ray pulsars, neutron stars (NSs) emitting in the X-ray band and showing coherent pulsations originated by the NS rotation around its axis. In particular, the MCs host an unusually high number of High Mass X-ray binaries (HMXBs), systems in which the compact object, a NS or a black hole (BH), orbits an early (spectral type O/B) companion.

BeXRBs (HMXBs with a bright Be spectral-type companion; see \citealt{Coe:2000,Okazaki:2001} for a review) represent the most numerous subclass of HMXBs in the MCs, with $\sim$150 BeXRBs out of $\sim$160 HMXBs. A total of $\sim$80 BeXRBs, in particular, host X-ray pulsars \citep{Coe:2015,Antoniou:2016,Haberl:2016,Haberl:2022}. 
%about half of the 121 known HMXBs in SMC are BeXRBs \citep{Coe:2015,Haberl:2016}, while \cite{Antoniou:2016} report 33 BeXRBs out of 40 HMXBs in the LMC, with more than two-thirds of all the BeXRBs hosting an accreting pulsar \citep{Haberl:2022}. 
%The lower number of known HMXBs in the LMC (despite being an order of magnitude more massive than the SMC) may originate from  %a combination of the galaxy larger extension, resulting in an incomplete coverage of the LMC 
%in the X-ray band outside the central region \citep{Vasilopoulos:2018}, and 
%the more recent star formation burst. 
%This would also explain why, even within the SMC itself, BeXRBs are scarcer in the Wing, where a star formation burst happened later, $\sim$11\,Myr ago \citep{Antoniou:2010}.
%a timescale like the one observed in the SMC is more consistent with the evolutionary timescale of the Be phenomenon. %\textbf{PE: unclear to me.} 
The recent discoveries of new HMXBs and pulsars in BeXRBs in the outskirts of the LMC suggest that more systems are likey hidden in these regions \citep{Vasilopoulos:2018,Maitra:2021,Haberl:2022,Maitra:2022}. 
Only one Low Mass X-ray Binary (LMXB, i.e. XRB whose companion has a mass $M\lesssim1\solarM$) hosting an accreting NS is known in the LMC, namely LMC X-2. The source is persistent and shows thermonuclear bursts, testifying to the presence of an old/low-magnetic field strength NS. No coherent or quasi-coherent periodic signals have been reported for LMC X-2 \citep[see][and reference therein]{Agrawal:2020}. 

The SMC and the LMC are also known to host two magnetars, CXOU\,J010043.1--721134 \citep{Lamb:2002,LambAddendum:2003,McGarry:2005,Tiengo:2008,Chatterjee:2021} and SGR\,0526--66 \citep{Mazets:1979,Kulkarni:2003,Tiengo:2009,Park:2020} respectively. Magnetars \citep{Duncan:1992} are isolated NSs whose emission is (in the majority of the cases) in excess of their spin-down luminosities. Their emission is thought to be  powered (for the most part) by the release of energy by the strong ($10^{14}\,\mathrm{G}\lesssim B\lesssim10^{16}$\,G) magnetic fields in their interior \cite[for recent reviews, see][]{Turolla:2015,Kaspi:2017,Esposito:2021}%(for recent reviews, see \citealt{Kaspi:2017,Turolla:2015})
. % Their spin periods are found to be clustered in the 0.1--12\,s range and are characterised by large spin-down rates ($\dot{P}$ in the $10^{-11}$--$10^{-13}$\,s\,s$^{-1}$ range for the majority of the population). 
%These sources are usually discovered due to the detection of peculiar, short (lasting 1--100\,ms) X-ray or $\gamma$-ray bursts and/or during X-ray active phases known as outbursts, when their persistent X-ray luminosity rapidly increases up to $\sim10^{36}\ergs$. 
%Magnetars represent the class of isolated X-ray pulsars that show the most extreme variability.
%, which can be a factor 1000 higher than their quiescent luminosity. 
Currently, about 30 magnetars are known, all in our Galaxy with the exception of the two magnetars in the MCs.  %Contrary to the case of BeXRBs, the optical/IR emission from magnetars is very faint (K$\sim$21$\magnitude$) and detected only for six magnetars in our Galaxy \cite[but see also][for an updated catalogue of magnetars with candidate optical/NIR counterparts]{Chrimes:2022}. %: 4U\,0142+61 \citep{Hulleman:2004,Dhillon:2005}, SGR\,0501+4516 \citep{Dhillon:2011}, 1E\,1048.1-5937 \citep{Wang:2002,Dhillon:2009}, SGR\,1806-20 \citep{Israel:2005}, XTE\,J1810-197 \citep{Israel:2004} and 1E\,2259+586 \citep{Hulleman:2001}. The optical/NIR component is thought to be related to the same physical mechanism responsible for the high-energy emission or, alternatively, to a surrounding disc. %Optical pulsations have been detected in the optical band in three magnetars \citep{Kern:2002,Dhillon:2005,Dhillon:2009,Dhillon:2011}.

In this paper, we report on the discovery of X-ray pulsations at a period of about 7.25\,s from a serendipitous source in the outskirts of the LMC, namely \belzlong\ (\belz\ henceforth). 
%, originally reported in the latest version of the \xmm\ serendipitous source catalogue 4XMM-DR12\footnote{\url{https://xcatdb.unistra.fr/4xmmdr12/xcatindex.html?detid=108416605010005}.} \citealt{Webb4XMM:2020}). 
Although an unambiguous classification is still missing, the discovery of the 7.25-s periodicity that we report in this paper is strongly suggestive of a spinning NS. 
%Although the vast majority of pulsars in the MCs are found in BeXRBs, 
We discuss two competing scenarios capable of accounting for the source's observed properties: an accreting NS hosted in a %HMXB
very rare evolutionary outcome for an XRB or an isolated NS belonging to the magnetar class. In the latter case, \belz\ would represent the third known magnetar in the MCs. 

The article is structured as follows: in Sect.\,\ref{sec:ObsDataReduction} we describe the observations analysed in this article and the data processing techniques that we applied. In Sect.\,\ref{sec:results}, we report on the results of our timing and spectral analyses. We discuss our results and the possible nature of this newly discovered pulsar in Sect.\,\ref{sec:Discussion}. Our conclusions are  in Sect.\,\ref{sec:conclusions}.

% This is a simple template for authors to write new MNRAS papers.
% See \texttt{mnras\_sample.tex} for a more complex example, and \texttt{mnras\_guide.tex}
% for a full user guide.

% All papers should start with an Introduction section, which sets the work
% in context, cites relevant earlier studies in the field by \citet{Fournier1901},
% and describes the problem the authors aim to solve \citep[e.g.][]{vanDijk1902}.
% Multiple citations can be joined in a simple way like \citet{deLaguarde1903, delaGuarde1904}.

\section{Observations and data reduction}\label{sec:ObsDataReduction}

\begin{figure*}
    \centering
    \includegraphics[scale=0.288]{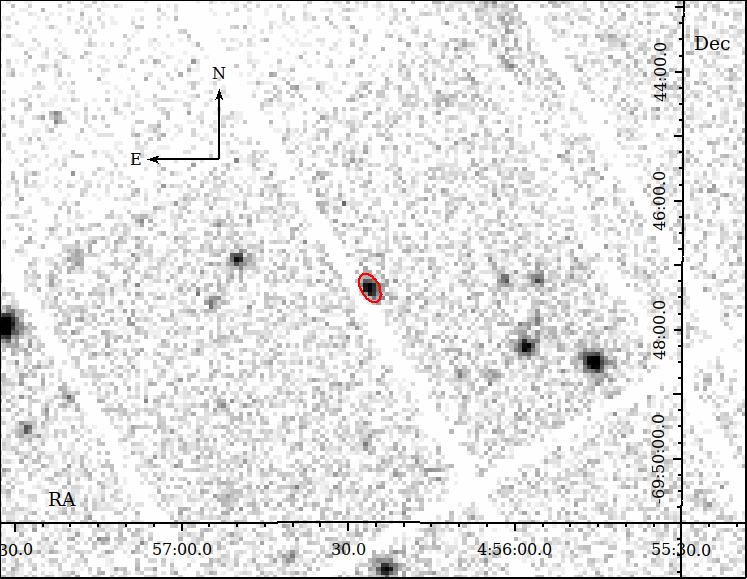}
    \includegraphics[scale=0.312]{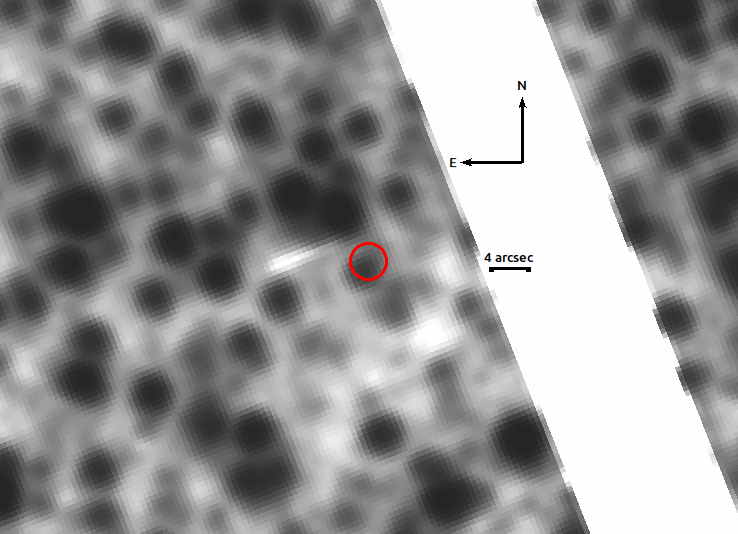}
    \caption{Close-up view of the \xmm\ (left panel) and \salt\ (right panel) field of view, obtained during obs. 0841660501 and the night of the 24 November, 2022, respectively. Note the difference in scale between the two images. The \xmm\ image shows both PN and MOSs data and the source region considered for the events extraction in red. The PN chip border is visible on the left of the source. The \salt\ image shows the \belz\ position with the associated error at the 3$\sigma$ level in red. The white gap on the right is due to the CCD border.% \paolo{Se non in scala (non è chiaro), non si potrebbe zoomare un più in XMM?}
    }
    \label{fig:SALTfield}
\end{figure*}

\subsection{X-ray observations}\label{sec:ObsDR_XMM}

\belz\ is located in an outer region of the LMC that has been observed twice by \xmm\ \citep{Jansen:2001}. The source was detected only during the second pointing (obs. 0841660501, duration $T\simeq47\,\mathrm{ks}$) in October 2019. EPIC PN \citep{Struder:2001} and EPIC MOS1 and MOS2 \citep{Turner:2001} data are available. The source was clearly detected in the three CCD cameras, which were operating in Full Frame mode (time resolution of 73.4\,ms for EPIC PN, 2.7\,s for EPIC MOS1 and 2.6\,s for EPIC MOS2). We used \textsc{sas} \citep{SAS:2004} v.20.0.0 and standard data reduction procedures to prepare the raw data for timing and spectral analysis. For the PN data, we considered only events with PATTERN $\le$ 4, while for MOS data we selected events with PATTERN $\le$ 12. Since \belz\ was observed near the border of a PN chip gap, we selected events for both timing and spectral analyses in a $14\times8\asec$ elliptical region aligned with the chip border and centred on the source position RA = $4^\mathrm{h} 56^\mathrm{m} 26\fs36$, Dec = $-69^\circ 47' 23\farcs1$. The background was evaluated using a large source-free circular region on the same chip with radius 80$\asec$. The same regions were considered for the MOS cameras. The event arrival times were converted to the barycentre of the Solar System using the \textsc{sas} task \texttt{BARYCEN} and the source position. %\textbf{PE: its unclear. Are we talking of the MOSs? (Same regions used for the PN?)} %rebinned to have at least 1 count per energy bin using the \textsc{sas} task \texttt{SPECGROUP}
Response matrices and ancillary files were produced using the \textsc{sas} tasks \texttt{RMFGEN} and \texttt{ARFGEN}, respectively. For our X-ray timing and spectral analyses we considered events in the 0.3--10\,keV band. Unless otherwise stated, the reported errors correspond to 1$\sigma$ (68.3\%) confidence ranges. A total of 776 events (source+background) were collected by the PN (426) and MOSs (350) cameras in the source extraction region, corresponding to a background-subtracted count rate of $(2.71\pm0.09)\E{-3}$\,count/s. 
% During the first visit (obs. 0823790301, length $T\simeq41\,\mathrm{ks}$) in 2018, we estimated a 3$\sigma$ upper limit\footnote{All the upper limits have been computed using the WebPIMMS tool: \url{https://heasarc.gsfc.nasa.gov/cgi-bin/Tools/w3pimms/w3pimms.pl}.} for the unabsorbed flux in the 0.3-10\,keV band $\FX=1.46\E{-14}\ergcms$. In the latter case we considered only MOS data since the source falls in the gap between two PN chips. A search  in the \chandra, \nustar\ and \swift\ archives gave negative results.
% %, but the source was always under the detection threshold. 
% \erosita\ \citep{Predhel:2021} observed the field of \belz\ during the all-sky surveys, providing the second most stringent 3$\sigma$ upper limit for the unabsorbed flux in the 0.3-10\,keV band $\FX=3.8\E{-14}\ergcms$. Two \rxte\ pointings (ObsID  40087-04-01-00 and 96441-01-01-00) were also searched for signals around the \xmm\ detected period without success.  
%{\color{red} Add RXTE data section here and in the results when Pg sends the reduced data...Analyzed obs: 40087-04-01-00 (obs with no jumps, shorte) and 96441-01-01-00 (obs with jumps, longer). Nothing to report...}

%We also checked the first \xmm\ visit (obs. 0823790301, length $T\simeq41\,\mathrm{ks}$) in 2018 considering only MOS data since the source falls in the gap between two PN chips, but the source flux was under the detection threshold. 
No pointings covering the source position are present in the \chandra\ and \nustar\ archives. 
20 \swift\ pointings are found, though the XRT throughput and the relatively short exposures provided poor constraints (see Sect.\,\ref{sec:SpectralResults} for more details). Two \rxte\ pointings were found and considered in the timing analysis (see Sect.\,\ref{sec:TimingResults}).  

%also searched for signals around the \xmm\ detected period without success. 

Finally, \erosita\ \citep{Predhel:2021} observed the field of \belz\ during the all-sky surveys but the source was always too faint to be detected.
Most of our analyses are based on the \xmm\ observation 0841660501 during which the source was detected. The left panel in Fig.~\ref{fig:SALTfield} shows a close-up view of \belz\ field as observed by \xmm\ during this observation, together with the extraction region of the source events (highlighted in red).

\subsection{Optical observations}\label{sec:ObsDR_SALT}

Following \cite{XMMcalibration:2004} and \cite{Webb4XMM:2020}, we adopted a 2$\asec$ uncertainty radius at 3$\sigma$ confidence level for the positional accuracy. %1.1$\asec$ uncertainty radius following the 4XMM-DR12 catalogue\footnote{\url{http://xmm-catalog.irap.omp.eu/source/208416605010005}.}
Within the \xmm\ uncertainty region of \belz\ we identified only one possible optical counterpart in the {\it Gaia} DR3 catalogue \citep{GaiaMission:2016,GaiaDR3:2022}, with a magnitude G$\sim$19.2$\magnitude$. %\textbf{PE: %Nominal distance? Gaia color?} 
We observed this source with the 11-m Southern African Large Telescope \salt\ \citep{Buckley:2006} during two consecutive nights, on November 24, 2022 and November 25, 2022. The right panel in Fig.~\ref{fig:SALTfield} shows a close-up view of \belz\ field as observed by \salt\ on November 24, 2022, together with the 3$\sigma$ uncertainty region of \belz\ (highlighted in red). The observations were divided in 5$\times$1200\,s exposures, for a total of 6000\,s. During both nights the seeing was $\sim$ 2$\asec$ with clear sky condition. We used the Robert Stobie Spectrograph RSS \citep{Burgh:2003,Kobulnicky:2003} in long-slit spectroscopy mode for a spectral classification of the source. %The RSS was set {\color{red} ask Gianluca how to continue, what I should include and so on} \emph{We assumed the following instrumental configuration: grating pg0700 plus order blocking filter pc03850, default camera station and grating angle. We considered the long-slit mode with a slit width of 1.5". The configuration was chosen to prevent the H$\alpha$ line from falling within the chip gaps.}, obtaining a spectral resolution of $\sim$735 at a central wavelength of 5581\,\AA. Wavelength calibration was performed with CuAr arc spectra. {\color{red} what else do we need? How we performed the data reduction? Waiting for David's response...} \\
The PG0700 grating was used with a tilt angle of 4.6$^\circ$, which resulted in a wavelength range of 3600 -- 7400\,\AA. The primary reductions, which include overscan corrections, bias subtraction, gain and amplifier cross-talk corrections, were performed with the \salt\ science pipeline \citep{Crawford:2012}. The remaining steps, including wavelength calibration, background subtraction and extraction of the one-dimensional spectra were performed with \textsc{iraf}. The spectrophotometric standard star, LTT1788, was observed with the same grating settings as our target with an exposure time of 60\,s. The reduction of the spectrum of LTT1788 was performed using the same procedure described above and a relative flux calibration was applied to our target spectrum.

\section{Results}\label{sec:results}

%For our X-ray timing and spectral analyses we considered events in the 0.3--10\,keV band. Unless otherwise stated, the reported errors indicate 1$\sigma$ (68.3\%) confidence ranges. A total of 776 events (source+background) were collected by the PN (426) and MOSs (350) cameras in the source extraction region, corresponding to a background-subtracted count rate of $(2.71\pm0.09)\E{-3}$\,count/s.
%the low number of events prevents a deeper analysis than the one reported in the following sections. 

    \subsection{Timing Analysis}\label{sec:TimingResults}
% {\color{blue}Analysis performed: timing analysis (insert PSD here), period estimated by the means of phase fitting (insert pulse profile, in the entire energy band and in subbands, bkg-subtracted). File per il timing riestratti usando la regione small (quella che userò per l'analisi spettrale). 
% \\

% %DA RIFARE: ERRORE NELL'ESTRAZIONE FILE. Frazione pulsata più alta...Alcuni valori sono stati già aggiornati, ricontrollare quando si inseriscono. 

% Per la stima della frazione pulsata usare i file in \texttt{all\_small\_lcbkgsub.lis}. $\mathrm{PF}=(85.6\pm3.7)\%$.
% Per il profilo in funzione dell'energia prendere come intervallo di energia 0.3-0.6 keV per la banda soft e 0.6-10. keV per la banda hard. Per banda soft $PF\sim79\%$, mentre per banda hard $PF\sim69\%$ (ma considerando gli errori sono compatibili a 1$\sigma$).
% \\

% Sembra esserci un secondo picco (in tutto l'intervallo energetico il fit migliora se aggiungo le altre armoniche, ma nelle singole bande l'evoluzione non è chiara).}
Pulsations from \belz\ were first detected within the framework of an extension of the Exploring the X-ray Transient and variable Sky (EXTraS) project\footnote{The details of the periodic signal-search pipeline can be found in Sect.\,4 of \cite{DeLuca:2021}.
%EXTraS was carried out from 2014 to 2016 and provided an unprecedented detailed description of the (a)periodic temporal variability of more than 400\,000 sources. See \url{https://cordis.europa.eu/project/id/607452}.
}  \citep[][Rodr\'{i}guez Castillo et al., in prep.]{DeLuca:2021}. 
We started from the EXTraS discovery parameters: a 
%We started our analysis from the results obtained by the EXTraS pipeline that detected a 
peak at a frequency $\nu\simeq0.138\,\mathrm{Hz}$ in the PN data with a confidence level of about 5.6$\sigma$.
%corresponding to a single PSD detection threshold at 3.5$\sigma$.  
%We tested the entire frequency range (from $\sim10^{-5}$\,Hz up to the Nyquist frequency $\nu_\mathrm{Nyq}\sim0.185$\,Hz). 
We followed \cite{Israel:1996} to infer the significance of the peak and the detection threshold in the PSD. As reported above, the vicinity of the source to the CCD gap motivated us to extract the events by using an elliptical region with the major axis aligned to the chip edge. The choice of an \textit{ad hoc} region for the extraction was also motivated by the fact that, given the peculiar position of the source, the EXTraS automatic pipeline failed to determine the best circular extraction region. %\textbf{PE: What does it mean `from scratch'? Are you explaining why you didn't rely on the EXTraS products?   } 
%The EXTraS pipeline extracts events in a circular region around the source by default. However, in the case of \belz, as described in the previous section, we extracted the source events in an \emph{ad hoc} region, since the default selection also included the chip gap. In Fig.~\ref{fig:belz_dps_PNMOSs} we report the 0.3--10\,keV power spectrum of the PN (top panel) and MOSs (bottom panel) data, together with the 3.5$\sigma$ detection threshold. 

\begin{figure}
    \includegraphics[width=\columnwidth]{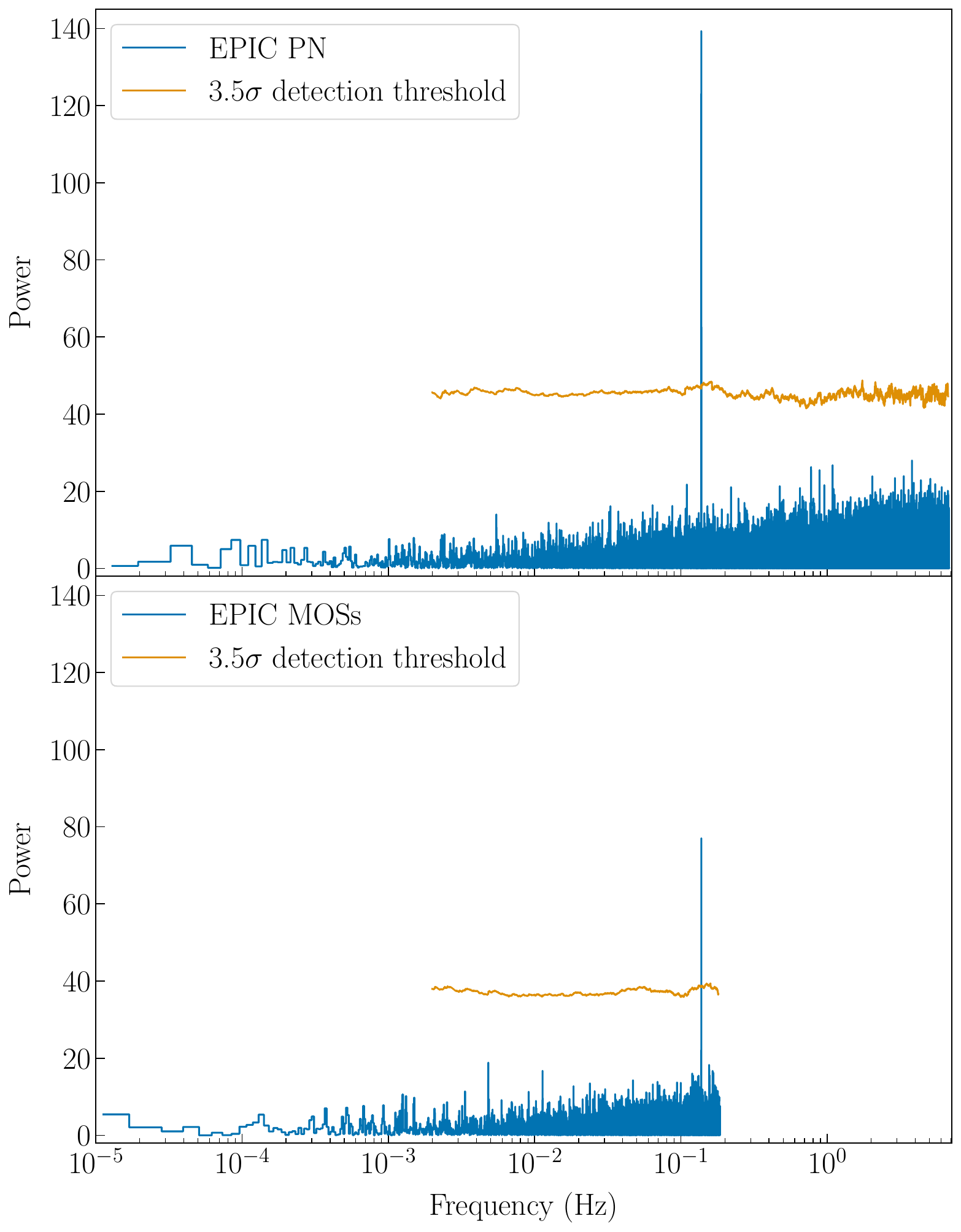}
    \caption{PSD (blue) of \belz\ light curve in the 0.3--10\,keV band obtained from EPIC PN (top panel) and MOSs (bottom panel) data of observation 0841660501, together with the local 3.5$\sigma$ detection threshold (orange). The peak at frequency $\nu\simeq0.138$\,Hz has a significance of 11.3$\sigma$ over the whole PN+MOSs PSD.}
    \label{fig:belz_dps_PNMOSs}
\end{figure}

In Fig.~\ref{fig:belz_dps_PNMOSs} we report the 0.3--10\,keV PSD of the PN (top panel) and MOSs (bottom panel) data, together with the 3.5$\sigma$ detection threshold. The peak is detected in the time series of the three EPIC instruments and not in the event lists of the other sources detected in the field of view. Moreover, no significant peaks are detected in the background regions extracted in different parts of the CCDs. Correspondingly, we can exclude that the detected signal is spurious or due to noise fluctuations. Within the combined PN+MOSs PSD the peak has a significance of 11.3$\sigma$ (computed over the whole $2\E{-5}-0.185\,\mathrm{Hz}$ PSD). %As expected from the white-noise dominated PSD, t
%The non-periodic variability of the lightcurve of the source during the observation is negligible
The PSD shows no sign of red-noise variability, with a 3$\sigma$ upper limit on the rms fractional variation of 0.36 from 10\,\textmu Hz up to 10\,mHz. 

% We refined the period estimate through epoch folding \citep{Leahy:1983} and phase-fitting techniques and obtained a value $P = 7.25243\pm0.00004\,\mathrm{s}$. The 0.3--10\,keV folded profile is reported in the top panel of Fig.~\ref{fig:efold_hardratio} (circle markers) and has a pulsed fraction (computed as the amplitude of the best-fit sinusoid) $PF=86.0\pm6.2\%$. We considered two energy sub-bands to study the evolution of the signal, a soft 0.3--2\,keV band and a hard 2--10\,keV band. The two sub-bands were chosen so that about half of the events would fall in each sub-band. In the top panel of Fig.~\ref{fig:efold_hardratio} we report also the folded profile in the soft (triangle markers) and hard (diamond markers) bands, where the pulsed fraction is equal to $PF_\mathrm{low}=94.9\pm7.0\%$ and $PF_\mathrm{hard}=84.8\pm9.5\%$, respectively.  

% EFOLD + HR CON COLORI DIVERSI
% \begin{figure}
%     \centering
%     \includegraphics[width=\columnwidth]{Immagini/belzebu_efolds_hardratio.pdf}
%     \caption{\emph{Top}: phase-folded profile of the PN+MOSs background-subtracted light curves in the 0.3-10 keV (blue circles), 0.3-0.6 keV (yellow triangles) and 0.6-10 keV (green diamonds) bands. The 0.3-0.6 keV and 0.6-10 keV light curves are vertically shifted with an offset equals to 1.5 and 3, respectively. \emph{Bottom}: hardness ratio between the two sub-bands.}
%     \label{fig:efold_hardratio}
% \end{figure}

% EFOLD * HR CON STESSO COLORE
\begin{figure}
    \centering
    \includegraphics[width=\columnwidth]{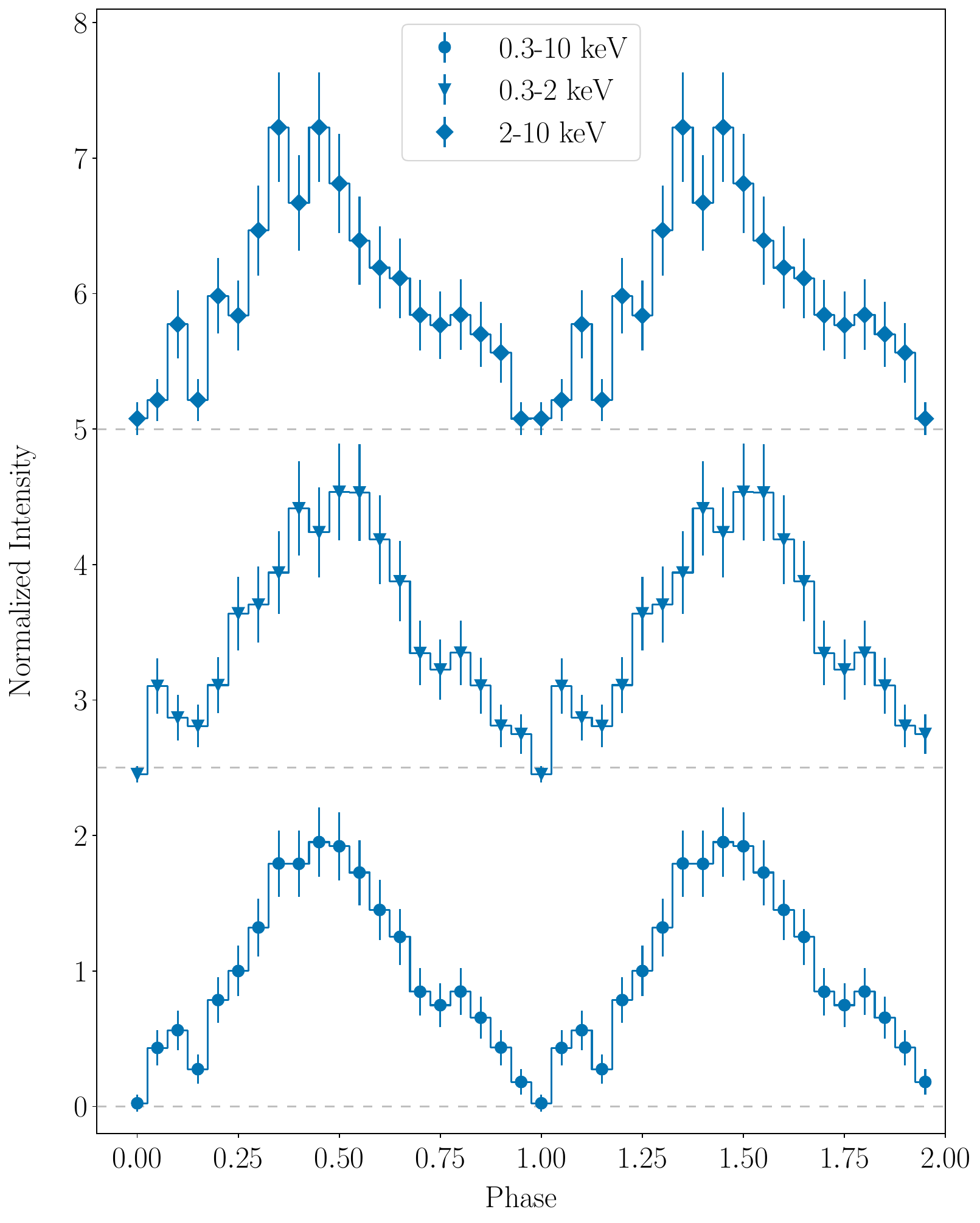}
    \caption{Phase-folded profile of the PN+MOSs background-subtracted light curves in the 0.3--10\,keV (circles), 0.3--2\,keV (triangles) and 2--10\,keV (diamonds) bands. For display purposes, the 0.3--2\,keV and 2--10\,keV light curves are vertically shifted with an offset equals to 2.5 and 5, respectively. The gray dashed lines show the zero-flux level for each energy band.% Bottom: hardness ratio hard/soft between the two sub-bands.
    }
    %\caption{From top to bottom: phase-folded profile of the PN+MOSs background-subtracted light curves in the 0.3--10\,keV, 0.3--2\,keV and 2--10\,keV bands, and hardness ratio hard/soft between the two sub-bands.}
    \label{fig:efold_hardratio}
\end{figure}

% EFOLD *HR STESSO COLORE NO LEGENDA
% \begin{figure}
%     \centering
%     \includegraphics[width=\columnwidth]{Immagini/belzebu_efolds_hardratio_onecolor_nolegend.pdf}
%     \caption{\emph{Top}: phase-folded profile of the PN+MOSs background-subtracted light curves in the 0.3-10 keV (circles), 0.3-0.6 keV (triangles) and 0.6-10 keV (diamonds) bands. The 0.3-0.6 keV and 0.6-10 keV light curves are vertically shifted with an offset equals to 1.5 and 3, respectively. \emph{Bottom}: hardness ratio between the two sub-bands.}
%     \label{fig:efold_hardratio}
% \end{figure}

We refined the period estimate through epoch folding \citep{Leahy:1983} and phase-fitting techniques and obtained a value of the period of $P_\mathrm{spin} = 7.25243\pm0.00004\,\mathrm{s}$. We derived a $3\sigma$ upper limit on its first derivative $|\dot{P}_\mathrm{spin}| < 2.1\E{-8}\,\mathrm{s}\,\mathrm{s}^{-1}$. The 0.3--10\,keV background-subtracted folded profile is shown in Fig.~\ref{fig:efold_hardratio} (bottom curve with filled circle markers) and has a pulsed fraction (defined as the semi-amplitude of the sinusoid divided by the source average count rate) of $PF=(86\pm6)\%$. No higher harmonics are needed to fit the profile well. %We did not include higher-order harmonics since we verified that they did not significantly improve our fit, with an improvement below 3$\sigma$. 
We also considered two sub-bands, 0.3--2\,keV  and  2--10\,keV (chosen so that about half of the events fall in each energy interval), to investigate the dependence of the signal shape as a function of energy. In Fig.~\ref{fig:efold_hardratio} we also report the folded profile in the soft (triangle markers, central curve) and hard (diamond markers, top curve) bands, where the pulsed fractions are %$PF_\mathrm{low}=(94.9_{-7.0}^{+5.1})\%$ %\textbf{[PE: I'd write $\mathbf{(94.9_{-7.0}^{+5.1})\%}$]} 
$PF_\mathrm{soft}>88\%$ ($1\sigma$ lower limit)
and $PF_\mathrm{hard}=(85\pm10)\%$, respectively. Even in these two cases, the inclusion of higher harmonics in the model did not significantly improve the fit.
%Although a second peak around phase 0.7-0.9 is visible, especially in the hard band and in the whole 0.3--10\,keV band, adding a first harmonic to the fit did not improve the fit significantly (an improvement below 3$\sigma$), therefore for each band we fitted the signal with a sinusoid. 
The hard and soft pulsed fractions are consistent, within 3$\sigma$, with each other and with the same, 100\% pulsed fraction. Note that all pulse minima are consistent with having zero counts. 
%Considering the whole 0.3--10\,keV band we can derive a 3$\sigma$ lower limit of the pulsed fraction equal to 63.3\%. Correspondingly, we did not detect any variation of the hardness ratio
%We cannot claim an evolution in the hardness ratio (bottom panel in Fig.~\ref{fig:efold_hardratio}).
%, whose evolution in phase is consistent with the ratio being constant. 

For our timing analysis we also considered two RXTE pointings (ObsID  40087-04-01-00 and 96441-01-01-00) during which the \belz\ position was covered. 
%Given the non-imaging nature of the detector, we could not consider these observations for our flux upper limit analysis. 
We searched for signals around the \xmm\ detected period (we adopted a maximum $|\dot{P}_\mathrm{spin}|<$10$^{-10}$\,s\,s$^{-1}$) in the Proportional Counter Array \citep[PCA,][]{Jahoda:2006} data, without success. No reliable upper limit on the pulsed fraction could be inferred, being the PCA a non-imaging instrument, in which the background contribution to the count rate could not be reliably quantified. 
%{\color{red} Aggiungere frase su RXTE}

%Inserire spettro di potenza ed efold con bkg sottratto. Vedere HENefold per grafico più carino. Valutare se inserire curva di luce con bkg sottratto dinamicamente. \cite{Israel:1996} per metodo calcolo soglia di detezione. $P = 7.25243(4)\,\mathrm{s}$.

    \subsection{Spectral Analysis}\label{sec:SpectralResults}

%    {\color{red} Cambiare valori ottenuti dall'analisi spettrale una volta usato \cite{DickeyHI:1990} per l'nH, come suggerito da Frank.}

% INTRO SE USO LA CSTAT
% We performed the spectral analysis of \belz\ using the \textsc{xspec} package \citep{Arnaud:1996} v.12.12.0g, included within the \textsc{heasoft}\footnote{\url{https://heasarc.gsfc.nasa.gov/docs/software/l\textsc{heasoft}/}.} distribution (v.6.29c). Given the paucity of collected photons, a detailed spectral analysis is beyond the scope of this work. For the same reason, we searched for the best-fit parameters by the means of the Cash statistics \citep{Cash:1979,Kaastra:2017}. The abundances and cross sections were set on those by \cite{Wilms:2000} and \cite{Verner:1996}, respectively. In this section errors indicate a 90\% confidence range. 

% INTRO SE RAGGRUPPO DI UN FATTORE X.
We performed the spectral analysis of \belz\ with the \textsc{xspec} package \citep{Arnaud:1996} v.12.12.0g, which is included in the \textsc{heasoft}\footnote{\url{https://heasarc.gsfc.nasa.gov/docs/software/lheasoft/}.} distribution (v.6.29c). We grouped by three the single bins (to cope with the original spectral resolution) and grouped the data so as to have at least 15 counts in a single spectral bin (to enable a reliable usage of the $\chi^2$ statistics). % and grouped by three the single bins (to cope with the original spectral resolution)
Different rebinning factors (e.g., 20 counts per bin at least) did not significantly alter the estimated values of the parameters, but we chose a 15-count configuration because it provided a lower uncertainty on the absorption column to the source. The abundances and cross sections were set to those of \cite{Wilms:2000} and \cite{Verner:1996}, respectively. In this section, the uncertainties indicate a 90\% confidence range.

% SE TENGO RAGGRUPPATI PER 25 CONTEGGI ALMENO (ORA È PER ALMENO 15)
We first considered a simple absorbed power-law (PL) model, with two absorption components to account for both the Galactic and the inter-galactic plus LMC absorption: in \textsc{xspec} syntax, this model corresponds to \texttt{tbabs*tbvarabs*powerlaw}. We set the first absorption component (\texttt{tbabs}) to the Galactic value\footnote{Computed by means of NASA's HEASARC $N_\mathrm{H}$ web calculator \url{https://heasarc.gsfc.nasa.gov/cgi-bin/Tools/w3nh/w3nh.pl}.} $N_\mathrm{H,gal}=8.91\E{20}\nH$ in the direction of the LMC \citep{DickeyHI:1990}. Following \cite{Haberl:2022,Haberl:2023} we set the elemental abundances of the LMC \texttt{tbvarabs} component at 0.49 times that of the Solar abundance, and allowed the column density to vary. The PL model provides an excellent description of the spectrum ($\chi^2_r\sim1$). The associated best-fit parameters are reported in Table~\ref{tab:spec_models}. 

We also considered a model with an additional black body (BB) component, since an absorbed BB+PL is often used to fit the spectra's of both magnetars and accreting X-ray sources. The addition of a BB component to the PL model did not significantly improve the fit, resulting in a by-chance probability (computed by means of the F-test) of $\simeq1\%$, which led us to discard the model. We also considered an absorbed BB+BB model, since it is often used to fit the magnetars' spectra \cite[see, e.g.,][]{IsraelSGR1935:2016}. Our BB+BB model also provides a good description of the observed spectra, with $\chi^2_r\sim0.89$, and the values for the temperatures $kT_\mathrm{BB}$ ($\simeq$0.3 and 0.9\,keV, respectively) and radii $R_\mathrm{BB}$ ($\simeq$2 and 0.4\,km, respectively) of both BBs are consistent with the ones expected from hotspots on a NS. However, we favored the PL model given its greater simplicity, since it involves one emission component instead of two while still providing a good description of the spectrum. Both the best-fit parameters of the BB+PL and BB+BB model can be found in Table~\ref{tab:spec_models}.
%This corresponds to a significance of the BB component below 3$\sigma$, which led us to discard the model. 
We also tested a single component BB model, but the fit is significantly worse, as shown by a large $\chi^2_r\sim$1.5. 

%Our estimates for the best-fit parameters for all the tested models are reported in Table~\ref{tab:spec_models}. %For both the BB+BB and BB+PL model the intrinsic $N_\mathrm{H}$ is not well constrained and we could only derive a 90\% upper limit of $1.2\E{21}\nH$ and $2.9\E{21}\nH$, respectively. Moreover,
%Nevertheless, we report our best-fit parameters for the BB model for the sake of our discussion in the next section. 

\begin{figure}
    \centering
    \includegraphics[width=\columnwidth]{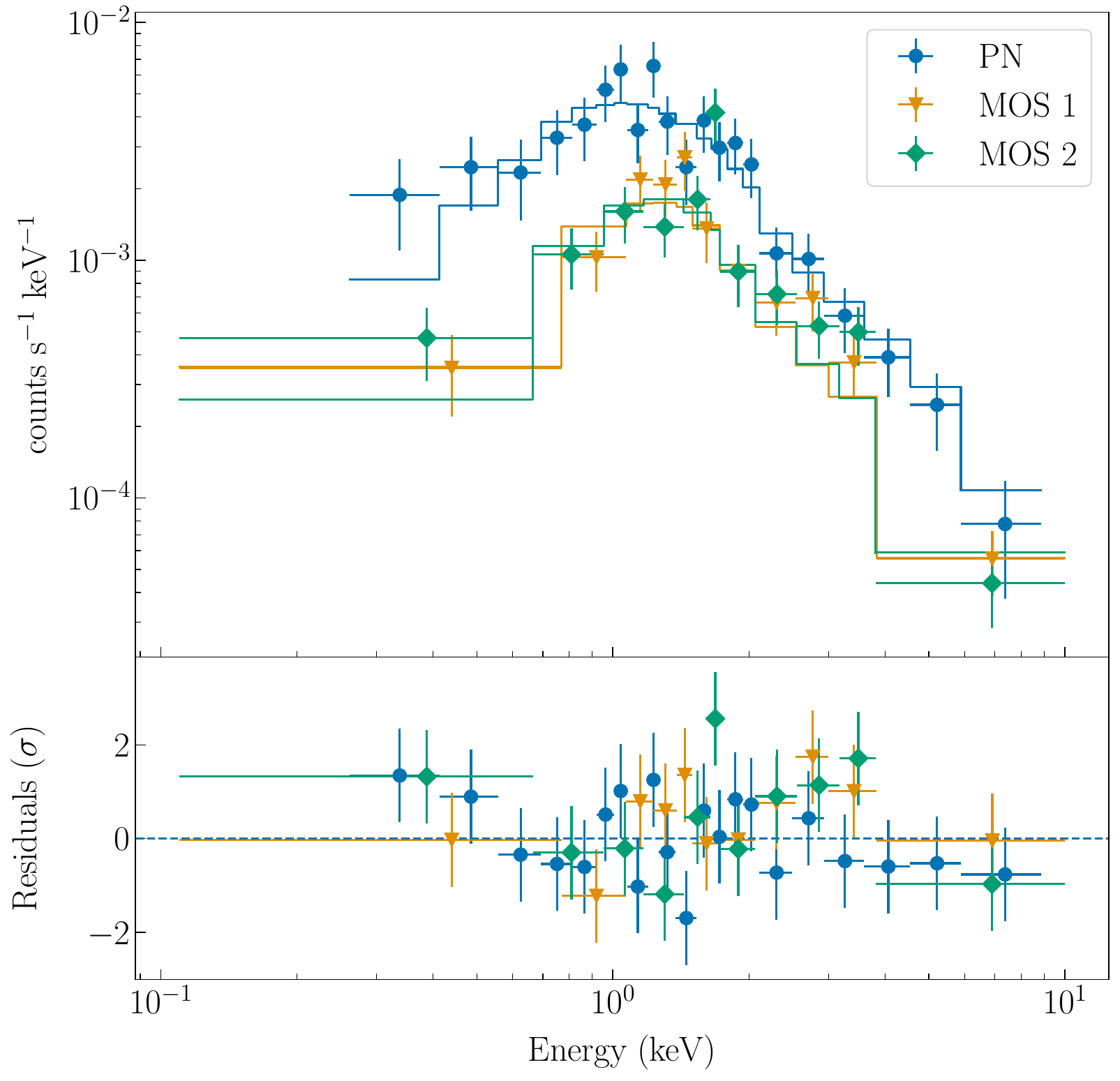}
    \caption{Top: EPIC spectrum of \belz\ from observation \obsid\ (blue circles: PN data; yellow triangles: MOS 1 data; green diamonds: MOS 2 data), together with the best-fit absorbed power-law model (same color scheme). Bottom: fit residuals in units of standard deviation (same color and marker scheme as above).}
    \label{fig:spectrum}
\end{figure}

The PN and MOSs spectra, together with the best-fit PL model, are shown in Fig.~\ref{fig:spectrum}. 
The intrinsic $N_\mathrm{H}\simeq4.8\E{21}\nH$ we derived by adopting the PL spectral model is consistent with the source being of extragalactic origin and the only known system along the line of sight is the LMC. Assuming a distance of 50\,kpc for the LMC \citep{LMCdistance:2013} and correcting for the total absorption, we derive an unabsorbed luminosity of about $\LX\simeq2.7\E{34}\ergs$ in the 0.3--10\,keV band. 

For those observations in which \belz\ was not detected we computed the 3$\sigma$ upper limit on its count rate by means of the \texttt{sosta} tool in \textsc{ximage}\footnote{\url{https://heasarc.gsfc.nasa.gov/xanadu/ximage/}.}. Exploiting the best-fit parameters reported in Table~\ref{tab:spec_models} for the PL model, we converted the 3$\sigma$ upper limits on the count rate in 3$\sigma$ upper limits on the unabsorbed flux of \belz\ in the 0.3--10\,keV band\footnote{All the upper limits have been computed using the WebPIMMS tool: \url{https://heasarc.gsfc.nasa.gov/cgi-bin/Tools/w3pimms/w3pimms.pl}.}. Most \swift\ pointings lasted less than 300\,s, too short to derive meaningful upper limits. From the longest observation (obs. 00033492002, $T\simeq630\,\mathrm{s}$, October 2014) we derived a 3$\sigma$ upper limit on the unabsorbed flux $\FX=9.2\E{-13}\ergcms$. The two most stringent upper limits come from the first \xmm\ pointing (obs. 0823790301, May 2018) and \erosita\ (May and November 2021, May and November 2022). %In the former case, by means of the \texttt{sosta} tool in \textsc{ximage}\footnote{\url{https://heasarc.gsfc.nasa.gov/xanadu/ximage/}.}, 
We derived a 3$\sigma$ upper limit on the count rate in the EPIC-MOS cameras of $9\E{-4}$\,counts/s, corresponding to a 3$\sigma$ upper limit on the unabsorbed flux $\FX=1.5\E{-14}\ergcms$. In the latter, considering the first four \erosita\ all-sky surveys (eRASS1-4), the derived 3$\sigma$ upper limit for the unabsorbed flux is $\FX=3.8\E{-14}\ergcms$ for each survey (F. Haberl, private communication). 

\begin{table*}
	\centering
	\caption{Best-fit parameters of the models considered for the phase-averaged spectrum of \belz. PL: power-law component. BB: black body component. $\Gamma$: power-law photon index. The reported errors show the 90\% confidence region. When only one black body component was adopted in the model we reported the temperature $kT_\mathrm{BB}$ and the radius $R_\mathrm{BB}$ under the cold ones column ($kT^\mathrm{cold}_\mathrm{BB}$ and $R^\mathrm{cold}_\mathrm{BB}$).}
	\label{tab:spec_models}
        \resizebox{\textwidth}{!}{
	\begin{tabular}{lccccccccr} 
		\hline
            \hline
		Model & $N_\mathrm{H}^a$ & $\Gamma$ & $kT^\mathrm{cold}_\mathrm{BB}$ & $R^\mathrm{cold}_\mathrm{BB}$ & $kT^\mathrm{hot}_\mathrm{BB}$ & $R^\mathrm{hot}_\mathrm{BB}$ & $F_\mathrm{X}^b$ & $L_\mathrm{X}^c$ & $\chi^{2}$/dof\\
            & $(10^{21}\nH)$ & & (keV) & (km) & (keV) & (km) & $(10^{-14}\ergcms)$ & $(10^{34}\ergs)$ & \\
		\hline
		PL & 4.8$^{+2.5}_{-2.0}$ & 1.93$^{+0.23}_{-0.21}$ & -- & -- & -- & -- & 6.10$^{+0.73}_{-0.71}$ & 2.73$^{+0.45}_{-0.33}$ & 40.03/40\\ \\
            %BB & $<0.21^d$ & -- & 0.637$\pm$0.055 & 0.82$^{+0.49}_{-0.41}$ & -- & -- & 4.45$^{+0.51}_{-0.49}$ & 1.43$\pm$0.15 & 60.17/40\\ \hline 
            BB + BB & $<1.2^d$ & -- & 0.30$^{+0.10}_{-0.09}$ & 1.9$^{+3.2}_{-1.5}$ & 0.86$^{+0.18}_{-0.14}$ & 0.43$^{+0.28}_{-0.40}$ & 5.27$^{+0.55}_{-0.66}$ & 1.58$^{+0.17}_{-0.20}$ & 33.68/38 
            \\ \\
		PL + BB & $<2.9^d$ & 1.4$^{+1.2}_{-0.5}$ & 0.63$^{+0.26}_{-0.12}$ & 0.63$^{+0.53}_{-0.56}$ & -- & -- &5.87$^{+0.84}_{-0.82}$ & 1.9$^{+1.0}_{-0.2}$ & 31.80/38\\ 
		\hline
            \hline
            \multicolumn{8}{l}{\textbf{Notes.}}\\
            \multicolumn{8}{l}{$^a$~The Galactic absorption component was fixed to $N_\mathrm{H,gal} = 8.91\E{20}\nH$ \citep{DickeyHI:1990}.} \\
            %Another absorption component, fixed to $N_\mathrm{H,gal} = 1.17\E{21}\nH$, was added to take into account the absorption given by the galactic medium.}\\
            \multicolumn{8}{l}{$^b$~Absorbed flux in the 0.3--10\,keV band.}\\
            \multicolumn{8}{l}{$^c$~Unabsorbed luminosity in the 0.3--10\,keV band, assuming a distance of 50\,kpc.}\\
            \multicolumn{8}{l}{$^d$~90\% confidence level upper limit.}\\
	\end{tabular}
        }
\end{table*}

    \subsection{Optical Analysis}

\begin{figure}
    \centering
    \includegraphics[width=\columnwidth]{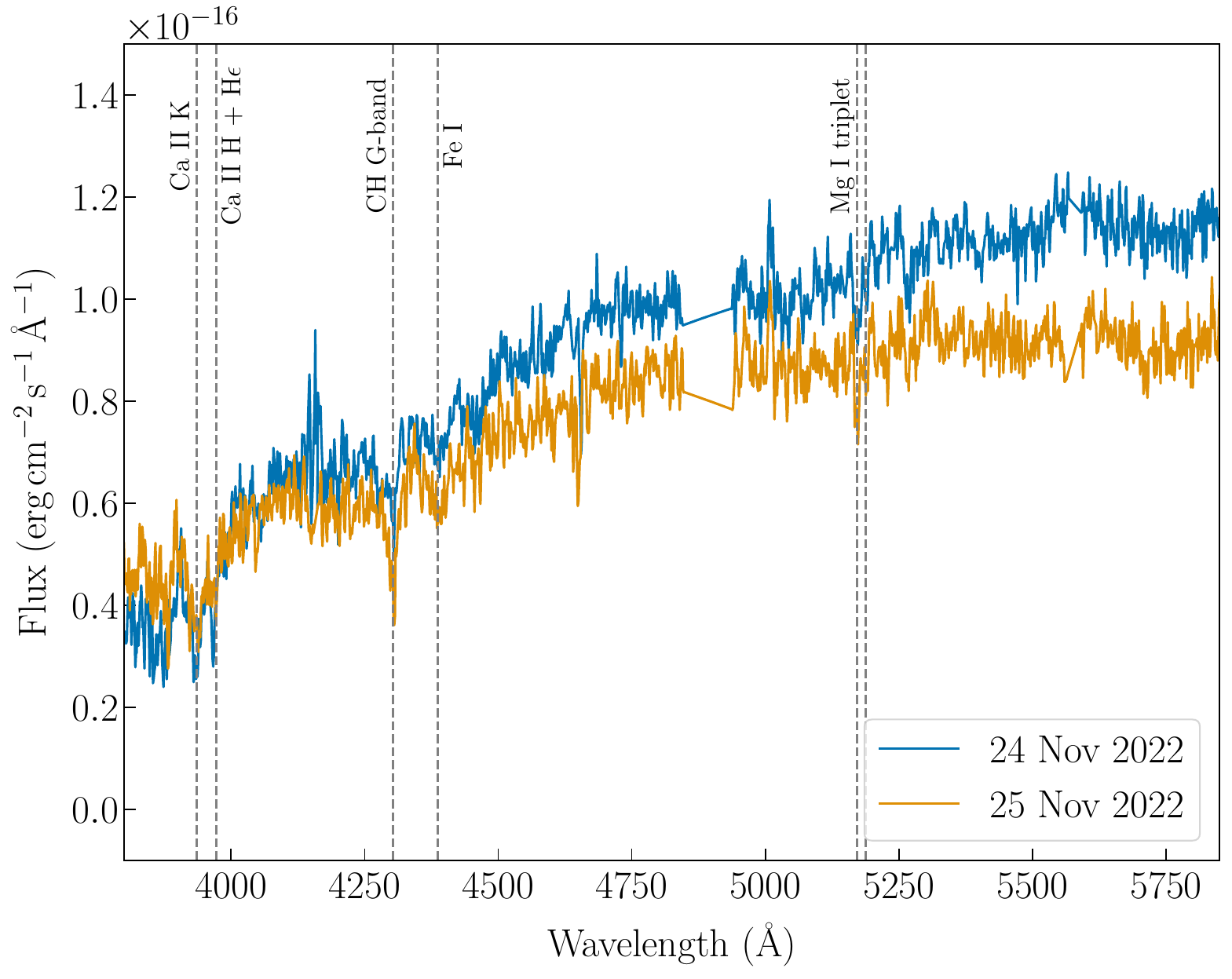}
    \caption{Optical spectra in the 3800--5850\,\AA\ band of the candidate optical counterpart of \belz\ during the first (blue) and second (yellow) night of observations. The features in the spectra we used for our classification are marked by the gray dashed lines.}
    \label{fig:SALTspectra}
\end{figure}

High-quality spectra were obtained in the 3800--5850\,\AA\ range for the optical object consistent with the X-ray position and are shown in Fig.~\ref{fig:SALTspectra}. The red-end of the spectrum (wavelength $\lambda > 5850$\,\AA) was dominated by diffuse emission of the surroundings during both nights. Therefore, we did not consider this range for our analysis. We derived a relative magnitude of $V\sim18.9\magnitude$. 

% \begin{figure*}
%     \centering
%     % \includegraphics[width=\columnwidth]{Immagini/LMCfield.png}
%     \includegraphics[scale=0.3]{Immagini/belzebu_xmm_border.png}
%     \includegraphics[scale=0.312]{Immagini/LMCfield_new.png}
%     \caption{Close-up view of the \salt\ image obtained during the night of the 24 November, 2022. \belz\ position with the associated error at the 3$\sigma$ level is highlighted in red. The white gap on the right is due to the CCD border.}
%     \label{fig:SALTfield}
% \end{figure*}

No emission lines usually associated with X-ray reprocessing by the companion and/or an accretion disc, such as the Balmer lines and He II (4686\,\AA), are found in the spectra. We first checked that the wavelengths of the few absorption lines present in the spectra are shifted by an amount equivalent to a radial velocity of $v_r\sim250\,\mathrm{km/s}$ ($\Delta\lambda\sim3.2-4.9$\,\AA\ in the 3800--5850\,\AA\ range), which is consistent with the star being in the LMC \citep{Piatti:2018}. Therefore, we associate this star with the LMC. Assuming a distance modulus $\mu\sim18.5\magnitude$ for the LMC \citep{Subramanian:2013} and an extinction coefficient\footnote{Computed with the NASA/IPAC Extragalactic Database (NED) Extinction Calculator available online at \url{https://ned.ipac.caltech.edu/extinction_calculator}.} $A_V\sim0.2\magnitude$ \citep{Schlafly:2011}, the absolute magnitude in $V$-band for this star is $M_V\sim0.2\magnitude$. This excludes any O or B spectral type star.
The optical spectra, in particular, show the Ca II K line at 3933\,\AA\ and the Ca II H + H$\epsilon$ blend at 3969\,\AA, together with the CH G-band at 4300\,\AA, the Fe I at 4383\,\AA\ and the Mg I triplet at 5167, 5172 and 5183\,\AA\ (all the absorption lines are reported to their restframe wavelengths). 
For our classification we considered spectral features that were visible in both observations. For instance, the lines at 4100--4150\AA, which appeared in emission during the first night (blue spectrum) and in absorption during the second night (yellow spectrum), were exluded as they probably arose from incorrect background subtraction. We can exclude that the source is a late-F star by the absence of lines such as H$\gamma$ (4340\,\AA), H8 (3889\,\AA) and H9 (3835\,\AA)% and Fe I (4325--4383\,\AA) lines
. The spectral shape, absolute magnitude $M_V$ and identified lines are all consistent with the source being a late-type G (G8) or early-type K (K0--3) star of luminosity class III. %From the Doppler shifts we measured for these lines we derive a radial velocity of $v_r\sim250\,\mathrm{km/s}$ ($\Delta\lambda\sim3.2-4.9$\,\AA\ in the 3800--5850\,\AA\ range), which is consistent with the LMC radial velocity \citep{Piatti:2018}. Therefore, we associate this star with the LMC. Assuming a distance modulus $\mu\sim18.5\magnitude$ for the LMC \citep{Subramanian:2013}, the absolute magnitude in $V$-band for this star is $M_V\sim0.4\magnitude$. 

% SPETTRO CON REBIN SOLO GRAFICO, FIT CON CSTAT
% \begin{figure}
%     \centering
%     \includegraphics[width=\columnwidth]{Immagini/belzebu_spectrum_min1.pdf}
%     \caption{\emph{Top}: EPIC spectrum of \belz\ from observation \obsid\ (blue circles: PN data; yellow triangles: MOS 1 data; green diamonds: MOS 2 data), together with the best-fit absorbed power-law model (same color scheme). \emph{Bottom}: fit residuals. In both panels data are rebinned for visual purpose only.}
%     \label{fig:spectrum}
% \end{figure}

% \begin{table}
% 	\centering
% 	\caption{Results of the spectral analysis of \belz. The analyzed model in \textsc{xspec} syntax is \texttt{tbabs*tbabs*powerlaw}.}
% 	\label{tab:spec_analysis}
% 	\begin{tabular}{lr}
% 		\hline
% 		Parameter & Value\\
% 		\hline
% 		1 & 2 & 3 & 4 & 5\\
% 		2 & 4 & 6 & 8 & 9\\
% 		3 & 5 & 7 & 9 & 10\\
% 		\hline
% 	\end{tabular}
% \end{table}

\section{Discussion}\label{sec:Discussion}

\begin{figure}
\centering
\includegraphics[width=\columnwidth]{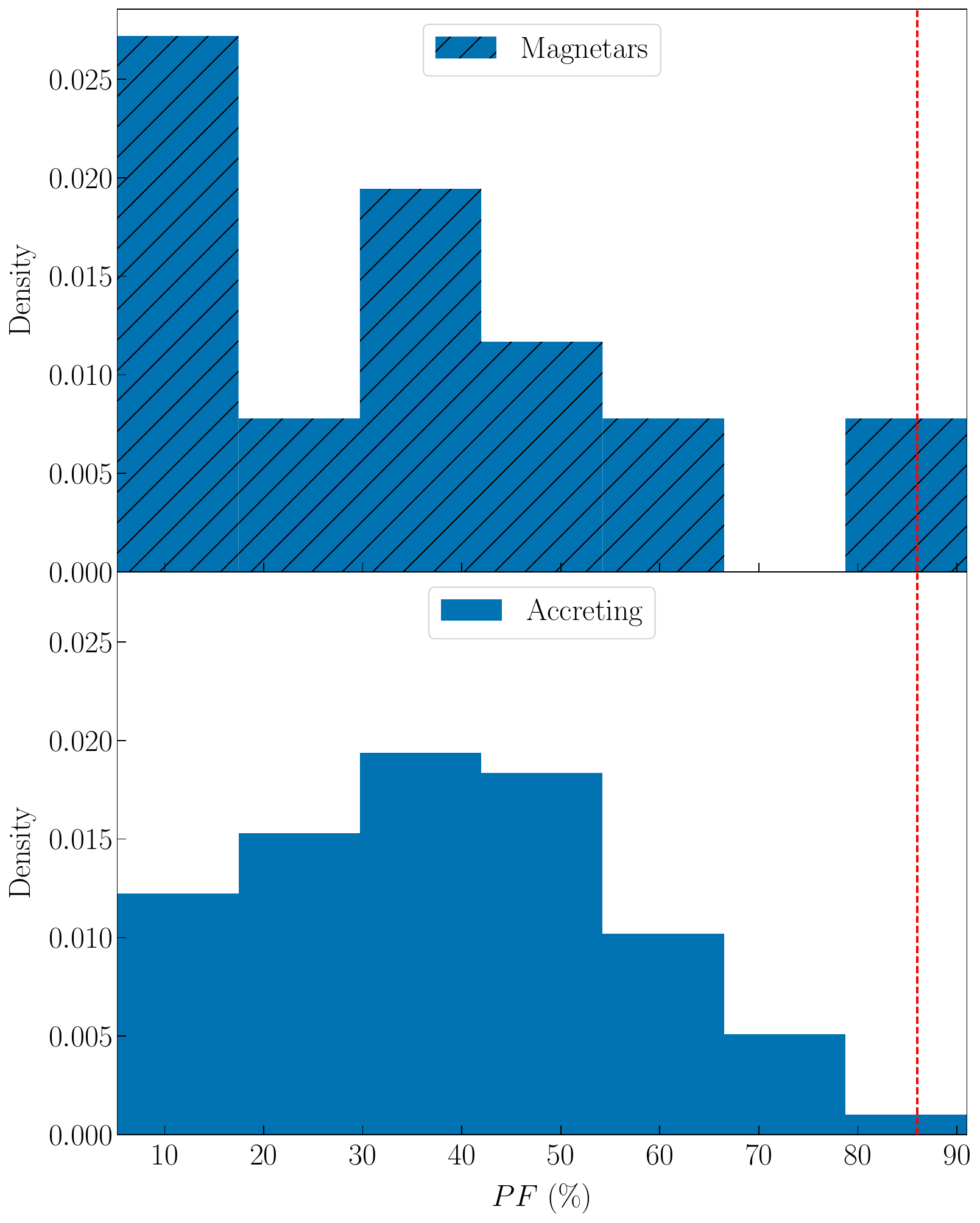}
\caption{
Top panel: distribution of the pulsed fraction of the magnetars reported in the MOOC catalogue \citep{CotiMOOC:2018} at the peak of the outburst .
Bottom panel: distribution of the pulsed fraction of the BeXRBs in the MCs during active phases \citep[and references therein]{Coe:2015,Antoniou:2016,Bartlett:2017,Boon:2017,HaberlF:2017,Kennea:2017,Vasilopoulos:2017,Koliopanos:2018,LaPalombara:2018,LaPalombaraSXP59:2018,Maitra:2018,Maitra:2021,Carpano:2022,HaberlSXP304:2022,Haberl:2022,Maitra:2022,Haberl:2023} and the pulsars in LMXBs Her\,X-1, GRO\,J1744-28, IGR\,J16358-4726, 3XMM\,J181923.7-170616 and Swift\,J1843.5-0343. For these last sources, we refer to the LMXBs XRBCat \citep{XRBCat:LMXB} and references therein. %To compute the luminosities, we assumed a distance of 50\,kpc \citep{LMCdistance:2013} and 62\,kpc \citep{Graczyk:2014} for the LMC and the SMC, respectively. 
In both panels the red dashed line marks the pulsed fraction of the signal found in \belz\ flux.
%Left panel: luminosity in the 0.3--10\,keV band in units of $10^{34}\,\mathrm{erg}\,\mathrm{s}^{-1}$ as a function of the spin period $P_\mathrm{spin}$. In this plot, we show J0456 (star marker), the magnetars (square markers) included in the MOOC catalogue \citep{CotiMOOC:2018}, the BeXRBs (triangle markers) in the Magellanic Clouds \citep[and references therein]{Coe:2015,Antoniou:2016,Bartlett:2017,Boon:2017,HaberlF:2017,Kennea:2017,Vasilopoulos:2017,Koliopanos:2018,LaPalombara:2018,LaPalombaraSXP59:2018,Maitra:2018,Maitra:2021,Carpano:2022,HaberlSXP304:2022,Haberl:2022,Maitra:2022,Haberl:2023} and the pulsars in LMXBs (circle markers) Her\,X-1, GRO\,J1744-28, IGR\,J16358-4726, 3XMM\,J181923.7-170616 and Swift\,J1843.5-0343. For these sources, we refer to the LMXBs XRBCat \citep{XRBCat:LMXB} and references therein. The color of each marker shows the pulsed fraction of the spin signal of the corresponding source, as indicated by the color bar in the middle. Right panel: distribution of the pulsed fractions of the magnetars (blue histogram with tilted lines) and the BeXRBS+LMXBs (orange histogram) considered in our work. The red dashed line marks the pulsed fraction of the signal found in J0456 flux. To compute the luminosities, we assumed a distance of 50\,kpc \citep{LMCdistance:2013} and 62\,kpc \citep{Graczyk:2014} for the LMC and the SMC, respectively.
}
\label{fig:comparisonhist}
\end{figure}

%{\color{red} Inserire in questa sezione la discussione sulla forma spettrale e quanto scritto nel commento riguardo gli spettri. Nel caso della forma del segnale il minimo a zero è consistente con un'autoeclissi, il che sarebbe difficile da spiegare nel contesto di una sorgente alimentata dall'accrescimento.}\footnote{\url{http://www.iasfbo.inaf.it/~mauro/pulsar_list.html}}

The 7.25-s coherent period of the pulsations strongly suggests that \belz\ is a spinning NS. 
% Indeed, the fastest-spinning accreting white dwarfs (WDs), i.e. LAMOST\,J024048.51+195226.9 and HD\,49798\footnote{Although the nature of HD\,49798 is still debated, a recent work by \cite{Mereghetti:2021} supports an accreting WD scenario.}, show a spin period of about 25 \citep{Pelisoli:2022} and 13\,s \citep{Israel:1997}, respectively.
%Although the short period of the pulsations is enough to classify \belz\ as a spinning NS (the shortest known spin period for an accreting white dwarf is about 13\,s; \citealt{Israel:1997}) {\color{red} Aggiungere commento sulla sorgente in \cite{Pelisoli:2022}, ma anche sul fatto che \cite{Mereghetti:2021} sembra confermare lo scenario WD per HD 49798}%\textbf{[PE: the fastest `certified' WD has a spin of 24.93 s . The nature of HD 49798, the 13-s one, si still somewhat debated.]}
However, the data currently available for \belz\ are not sufficient for an unambiguous classification. In fact, without a further measurement of the spin period $P_\mathrm{spin}$ it is not possible to derive its secular first derivative $\dot{P}_\mathrm{spin}$ and therefore to determine whether the pulsar is isolated (spin-down) or in an accreting binary system (spin-up).
%(a value that can help discern among the different classes of X-ray emitting NSs) and the derived upper limit on the intrinsic (intra-observation) $\dot{P}_\mathrm{spin}$ reported in Sect.~\ref{sec:TimingResults} is not meaningful. 
Therefore, we attempted a classification of the source based on the available information from the observations we analysed in the previous section and the MCs environment.
%All the findings can be summarized with \belz\ belonging to one of the following scenarios: (i) an accreting X-ray pulsar with a G8-K3 III companion or (ii) an isolated NS of the magnetar class. 
\belz\ may be identified as either (i) an accreting X-ray pulsar with a G8-K3 III companion or (ii) an isolated NS of the magnetar class.

%Among the HMXBs, the BeXRBs are known to show particularly high variability in their X-ray flux: \cite{Haberl:2016} studied the flux variability of BeXRBs in the SMC and found that the ratios of the maximum and minimum fluxes detected from these sources can be as high as $10^5$ (see their Fig.~5) at the period timescale of \belz. %{\color{red} aggiungere commento su Corbet diagram \citep{Corbet:1986}, a 7 s di spin mi aspetto un periodo orbitale di 10--30 d}%\textbf{[PE: maybe it's worth noticing that from the Corbet diagram, one would expect a 10--30 d orbital period, given the 7-s spin.]}
%Such a behaviour is associated to the presence of a decretion disc surrounding the Be companion: when the NS crosses the decretion disc, the NS is able to accrete more matter than usual and the X-ray luminosity rises accordingly. {\color{red} dipende anche dal fatto che il raggio di corotazione e' piu' piccolo e quindi ho bisogno di piu' Mdot per avere accrescimento sulla superfici}. The vast majority of X-ray pulsars in the MCs are found in HMXBs, so it is reasonable to assume that \belz\  could indeed represent the latest addition to this class. Considering the Corbet diagram \citep{Corbet:1986}, for an accreting NS with $P_\mathrm{spin}\simeq7\,\mathrm{s}$ whose companion is a Be star we expect an orbital period of 10--30\,d, though the sparsness of the available pointings does not allow us to test the. 

Concerning the first scenario, we note that we do not have any information on the X-ray activity state of the source at (or around) the time of the optical observations of \belz. Therefore, the absence of the optical emission lines, characteristic of X-ray reprocessing, does not rule out the accreting scenario,  as \belz\ could be a transient binary system in quiescence at the time of the \salt\ observations.

Given that no LMXB X-ray pulsars are known in the MCs, we started considering the properties of the known X-ray pulsars in the MCs, i.e. those in HMXBs, under the hypothesis that they are coeval with \belz. 
\cite{Christodoulou:2016} presented an analysis on all accreting X-ray pulsars with $P_\mathrm{spin}<10^3\,\mathrm{s}$ in both the LMC and the SMC. By looking at the relation between the minimum X-ray luminosity $\LX$ shown by these sources and their spin periods $P_\mathrm{spin}$ and following \cite{Stella:1986}, they were able to derive a typical lower limit on the magnetic field $B>3\E{11}\,\mathrm{G}$ for NSs with $P_\mathrm{spin} < 15\,\mathrm{s}$. Assuming that \belz\ is an accreting NS we can use the standard expression to derive the corotation radius $R_\mathrm{co}$
\begin{equation}
    R_\mathrm{co}=\left(\frac{GM_\mathrm{NS}}{\Omega^2}\right)^{1/3}\simeq1.5\E{8}\left(\frac{M_\mathrm{NS}}{M_\odot}\right)^{1/3}P^{2/3}\,\mathrm{cm}
\end{equation}
and %, following \cite{Campana:2018}, 
the magnetospheric radius $R_\mathrm{m}$
\begin{equation}
    R_\mathrm{m}=3.3\E{7}\xi_{0.5}B_{12}^{4/7}
L_{39}^{-2/7}R_6^{10/7}M_{1.4}^{1/7}\,\mathrm{cm,}
\end{equation}
where the X-ray luminosity $L$, the magnetic field $B$, the radius $R$ of the NS and its mass $M$ are in $10^{39}\ergs$, $10^{12}\,\mathrm{G}$, $10^6\,\mathrm{cm}$ and 1.4$\solarM$ units, respectively.
%\begin{equation}
%R_\mathrm{m}=3.3\E{8}\left(\mu_{30}^4L_{36}^{-2}R_6^{-2}\frac{M}{M_\odot}\right)^{1/7}\,\mathrm{cm}
%\end{equation}
%where $\mu_30$ is the magnetic dipole moment $\mu=BR^3$ in units of $10^{30}$\,G\,cm$^{3}$, $L_{36}=\LX\E{-36}\ergs$ and $R_6=R\E{-6}$\,cm. 
In our case $P\simeq7.25\,\mathrm{s}$, $\LX\simeq2.7\E{34}\ergs$. Assuming a standard value for $M_\mathrm{NS}\simeq1.4\solarM$, we derived $R_\mathrm{co}\simeq 6.3\E{8}\,\mathrm{cm}$. In order to have accretion onto the NS $R_\mathrm{m} < R_\mathrm{co}$. Assuming $\xi=0.5$, we find that the magnetic field of \belz\ should be $B < 9.1\E{11}\,\mathrm{G}$. Moreover, assuming that the non detection of the pulsar at a luminosity level a factor of about 4 lower ($\LX\simeq6.7\E{33}\ergs$) is related to the onset of the propeller phase, it converts to a magnetic field lower limit of $B > 4.5\E{11}\,\mathrm{G}$. Both limiting values are consistent with those obtained by \cite{Christodoulou:2016}, suggesting that the assumed coeval hypothesis is likely reliable.

%\cite{Zaritsky:2004}, however, lists only objects fainter then $V\simeq19\magnitude$ within 5$\asec$ from \belz\ position, while the companions of the known pulsars in BeXRBs in the LMC are all brighter than $V\simeq16.5\magnitude$. Moreover, the brightest star within this region shows colours which are inconsistent with an early-type star. 
%\salt\ observations clearly showed that the spectrum of the possible counterpart is consistent with that of a G/K-type star. Therefore, we cannot exclude that this star is indeed associated with \belz. The lack of emission lines in the optical spectrum cannot be considered an argument against the accretion scenario, since emission lines are related to the X-ray activity phase of the accreting object and \belz\ was not detected in November 2022 by \erosita\ (close to the optical observation epochs). However, an accreting X-ray pulsar with such a companion has not been observed yet in the MCs. 
Only two LMXBs are known in the MCs. eRASSt\,J040515.6-745202 \citep{HaberlLMXB:2023} showed a Type-I X-ray burst, indicative of low ($B\lesssim10^8$\,G) magnetic fields, and is not pulsating. LMC X-2, likewise non-pulsating, shows a blue optical spectrum and it is a persistent source \citep[see][and references therein]{Lavagetto:2008,Agrawal:2009,Agrawal:2020}.
Few Galactic LMXBs, such as Her\,X-1, GRO\,J1744-28, IGR\,J16358-4726, 3XMM\,J181923.7-170616 and Swift\,J1843.5-0343, host X-ray pulsars with similarly high ($B\gtrsim10^{11}\,\mathrm{G}$) magnetic fields, though they represent an exception among LMXBs. In fact, the greatest part of known pulsars in LMXBs are usually old NSs spun-up to ms-long spin periods by accretion and, therefore, possess low magnetic fields $B\lesssim10^8\,\mathrm{G}$ \citep{Bahramian:2022}, inconsistent with our findings.
%Pulsars in LMXBs are usually old NSs spun-up to ms-long spin periods by accretion and therefore possess low magnetic fields $B\lesssim10^8\,\mathrm{G}$ \citep{Bahramian:2022}  inconsistent with our findings. There are a few galactic X-ray pulsars in LMXBs which possess high ($B\gtrsim10^{11}\,\mathrm{G}$) magnetic fields, but these systems (such as Her\,X-1, GRO\,J1744-28, etc.) represent exceptions. 
Furthermore, the latter class of LMXBs very often shows aperiodic variability in the form of non-Poissonian power spectrum noise components (likely originated by the accretion process itself). The latter is simply not detected in the \xmm\ dataset  (see the PSD shown in Fig.~\ref{fig:belz_dps_PNMOSs}). Note that, following the recently released LMXB Cat\footnote{\url{http://astro.uni-tuebingen.de/~xrbcat/} and references therein.} \citep{XRBCat:LMXB}, for our discussion we included the symbiotic X-ray binaries (SyXBs, XRBs composed of an accreting NS and a late-type K1-M8 companion; see Tab.~1 in \citealt{Yungelson:2019} for a list of confirmed and candidates SyXBs) among the LMXBs. Summarizing, within the accreting X-ray pulsar scenario, \belz\ might be a new member of the rare class of relatively young pulsars in LMXBs/SyXBs, the first ever in the MCs, with very peculiar properties. 

The XRB scenario is also challenged by a previous work of \cite{Antoniou:2016}. Starting from the LMC star formation history reconstructed by \cite{Harris:2009} and the data coming from the Magellanic Clouds Photometric Survey \citep[MCPS, ][]{Zaritsky:2004}, they estimated the chance coincidence probability for an optical source to be found in the uncertainty region of an X-ray source in the LMC as a function of both the absolute magnitude $M_V$ and the reddening-corrected color index $B-V$. As one can see from Fig.~2 in the original paper, for a source with $M_V\sim0.2\magnitude$ like ours this probability is always $\gtrsim62\%$ within a $5\asec$ region. Their results are probably (at least partially) driven by the fact that they included the central region of the LMC, where more massive (O/B-type) stars are expected. Even considering this effect and the fact that they considered a bigger uncertainty region than us, we expect this probability to be only a factor of few higher than the chance coincidence probaiblity for \belz, especially considering how crowded the \salt\ field is (as can be seen in Fig.~\ref{fig:SALTfield}).

It is worth noticing that a period of 7.25-s 
might be still compatible with the spin period of an 
%by a we cannot totally exclude the possibility that \belz\ is instead an 
accreting white dwarf (WD). 
%However, several fact led us to dicard this scenario. 
For example, the fastest-spinning accreting WDs, i.e. AE Aqr, LAMOST\,J024048.51+195226.9 (though X-ray quiet) and HD\,49798\footnote{Although the nature of HD\,49798 is still debated, a recent work by \cite{Mereghetti:2021} supports an accreting WD scenario.}, show a spin period of about 33 \citep{LiAEAqr:2016}, 25 \citep{Pelisoli:2022} and 13\,s \citep{Israel:1997}, respectively. Correspondingly, in the accreting WD scenario \belz\ might be associated to an intermediate polar (IP) systems similar to those above cited.
If so, \belz\ would be simultaneously the fastest and brightest (typical luminosities\footnote{See the IP catalog \url{https://asd.gsfc.nasa.gov/Koji.Mukai/iphome/catalog/alpha.html} and references therein.} are $\LX\lesssim10^{32} - 10^{33}\ergs$) accreting WD of the whole IP class, the first ever in the MCs. The lack of orbital modulation in the spin signal phase is also an indication that, if \belz\ were an accreting source, the orbital period should be at least 3--4 times the duration of the observation ($T\simeq47$\,ks). This corresponds to $P_\mathrm{orb}\gtrsim40$\,h, the second longest orbital period for an IP after GK\,Per \citep[with an orbital period of $\sim$2\,d,][]{GKPersPorb:2021}, if not longer. Finally, IPs often show strong optical emission lines both during quiescence and active phases \citep{Saito:2010}. Our \salt\ spectra, on the other hand, do not show similar features. On the light of these factors, we consider this scenario rather unlikely. We also took into account the possibility of an isolated (non-accreting) WD like AR Sco \citep{Takata:2018}. However, \belz\ is 4 orders of magnitude brighter than AR Sco and the alternation active-quiescent phases is not expected in isolated sources. Therefore, we consider this scenario unlikely too.

If \belz\ is not associated with the G8-K3 III star, it might be related to the other class of variable NSs observed in the MCs, e.g. magnetars. %, %In this case, \belz\ might be a new member of the magnetar class. %The variability shown by \belz\ in the two \xmm\ observations is in fact typical of transient magnetars, though in this case it would mean that we have missed the onset of the outburst. 
%We also lack a measure of the $\dot{P}_\mathrm{spin}$, which in the case of magnetars is particularly high. %\textbf{[PE: Any meaningful upper limit? On the spin-up in particular, since a BeXRBs in high state can experience very intense spin up: the CATSBAR in the SMC had P/Pdot $<$ 100 yr.]}
%{\color{blue} The variability in flux shown by \belz\ in the two available \xmm\ observations is typical of transient magnetars, but in this case it means that for this source we missed the onset of the outburst. A $\dot{P}\sim10^{-15}-10^{-10}\pdotunit$ would be a clear indication that the source is indeed a magnetar, but the analysed observation is too short to constraint this value...the non-detection of Balmer lines in the optical candidate counterpart cannot allow us to exclude that the source is a BeXRB (only with a detection we could have claimed that)...} 
%there are several pieces of evidence that support the hypothesis that \belz\ is a magnetar in the LMC. 
In order to further explore this scenario 
we compared the source X-ray luminosity $\LX$ in the 0.3--10\,keV band and the spin period $P_\mathrm{spin}$ with the ones shown by all the magnetars included in the Magnetar Outburst Online Catalogue\footnote{An online and updated version of this catalogue can be found in the Magnetar Outburst Online Catalogue (MOOC): \url{http://magnetars.ice.csic.es/\#/welcome}. A catalogue of known magnetars has been compiled also by \cite{McGill:catalogue} and an updated version can be found at \url{http://www.physics.mcgill.ca/~pulsar/magnetar/main.html}.} \citep{CotiMOOC:2018}. We also compared them with the same quantities shown by the BeXRBs in the SMC and the LMC for which both $\LX$ and $P_\mathrm{spin}$ measurements were available during active phases. 
%for which the $\LX$ in the 0.3--10\,keV band, the spin period $P_\mathrm{spin}$ and the pulsed fraction $PF$ were computed/measured or could be derived with the available information in the literature. 
We considered the BeXRBs reported by \cite{Coe:2015} for the SMC and \cite{Antoniou:2016} for the LMC. We then cross-checked our sample in the Bologna INAF/IAPS catalogue of accreting X-ray binaries\footnote{\url{http://www.iasfbo.inaf.it/~mauro/pulsar_list.html} and references therein.} to check for missing pulsars not included in the aforementioned catalogues/publications.  To convert the flux in luminosity we considered a distance of 50\,kpc \citep{LMCdistance:2013} and 62\,kpc \citep{Graczyk:2014} for the LMC and SMC, respectively. We also included the same quantities for the few Galactic high-B X-ray pulsars found in LMXBs. Since \belz\ has been detected only once, we probably observed the source during a period of high activity. In order to compare these sources with \belz, we considered their $\LX$ and $P_\mathrm{spin}$ values as inferred during their outburst peak (in the case of magnetars) or during their high luminosity phases (in the case of BeXRBs and LMXBs). We considered the persistent luminosity values for those magnetars for which we do not have detection of an outburst: 1E\,1841-045 \citep{Vasisht:1997}, SGR\,1900+14 \citep{Mereghetti:2006}, 4U\,0142+614 \citep{ReaMNRAS:2007}, 1RXS\,J170849.0-490910 \citep{Rea:2007}, CXOU\,J010043.1-721134 \citep{McGarry:2005}, CXOU\,J171405.7-381031 \citep{Gotthelf:2019} and SGR\,0526-66 \citep{Mazets:1979}. 
%Although the latter is the first magnetar to have shown a Giant Flare back in 1979, the source has not experienced an outburst since then. Therefore, we considered this magnetar as quiescent since the values we considered were estimated $\approx$30 years later by \cite{Tiengo:2009}. 
Finally, for the LMXBs (often showing large flux ranges) we assumed average values of the fluxes, while we assumed the mean distance reported by the LMXB Cat %\footnote{\url{http://astro.uni-tuebingen.de/~xrbcat/} and references therein.} 
to convert the flux in luminosity% \matt{Maybe add here a comment about SyXRBs, to clearly state that we consider them among the LMXBs}
.

\belz\ is a peculiar case where both a short $P_\mathrm{spin}$ and low $\LX$ are observed, a combination typically associated with magnetars only. Indeed, the magnetars in our sample are found to have a luminosity $\LX\simeq10^{34}-10^{36}\ergs$. Accreting X-ray pulsars at similar ($\sim$7\,s) periods, instead, show a luminosity which is at least one order of magnitude higher ($\sim10^{35}\ergs$), with the least luminous XRBs at these periods in the MCs being LXP 8.04 at $\LX\simeq1.5\E{35}\ergs$ \citep{Vasilopoulos:2014} and SXP 7.92 at $\LX\simeq1\E{36}\ergs$ \citep{Bartlett:2017}.

%A self-eclipsing hotspot on an isolated NS, instead, is a more natural explanation. 
%{\color{blue} In the case of an accreting source it is hard to explain that the minimum of the signal is consistent with zero flux. More similar to an auto-eclipse of the hotspot from the source, hard to explain in the context of an accreting source...} %Nel caso della forma del segnale il minimo a zero è consistente con un'autoeclissi, il che sarebbe difficile da spiegare nel contesto di una sorgente alimentata dall'accrescimento.
A second indication of the magnetar nature of \belz\ comes from spectral analysis. As already explained in Sect.~\ref{sec:SpectralResults}, a simple power-law (PL) model was enough to obtain a good fit of the energy spectrum. Although both accreting X-ray pulsars and magnetars can exhibit a power-law component in their spectra, a photon index $\Gamma\simeq1.9$ is particularly high for an accreting system. A photon index $\Gamma\sim2$, instead, is in line (although not particularly high) with the values usually found for transient magnetars \citep{Esposito:2021}. 
It is also interesting to note that in the BB+BB model, often employed to describe magnetars' emission in outburst, the radii and temperatures associated with the BBs would be $R\simeq2$\,km and 0.4\,km and $kT_\mathrm{BB}\simeq0.3$\,keV and 0.9\,keV, respectively. Both values are compatible with those found for other magnetars \cite[see][and references therein]{Esposito:2021}. In this magnetar scenario, the dipolar magnetic field is expected to be of the order of 10$^{14}$ G.

We further note that high-amplitude pulsed fractions are more common in magnetars than in accreting pulsars. Fig.~\ref{fig:comparisonhist} shows the distribution of $PF$s for the magnetars (blue barred histogram in the top panel) and BeXRBs+LMXBs (blue histogram in the bottom panel) in our sample compared to the $PF$ inferred for \belz\ (red dashed line). For the $PF$ we considered the value shown during an outburst/high-active phase as for $\LX$ and $P_\mathrm{spin}$. 
% The XRBs' $PF$ distribution peaks at $PF\sim30-40\%$ and rapidly decays at higher $PF$s, while magnetars show a flatter distribution. 
At $PF>80\%$ (the range of interest for \belz) the probability of finding an XRB with properties similar to those of \belz\ is rather low, while it is not negligible for magnetars. Indeed, 2 out of the 21 magnetars in our sample show a pulsed fraction $PF>80\%$, while only one XRB, SXP\,95.2 \citep{Laycock:2002}, out of 80 shows such a high pulsed fraction. We also note that \belz\ pulse profiles in Fig.~\ref{fig:efold_hardratio} show a minimum consistent with a null flux, which is hard to explain in the framework of an accreting X-ray source, where low amplitude modulations are expected due to the presence of the accretion disc. There are no X-ray pulsars in the MCs with average pulse minima consistent with a zero flux level. In the case of isolated NSs, on the other hand, a zero-flux level can be easily explained in terms of geometrical occultation of an emitting region in the proximity of the surface by the NS itself. 

Finally, the ($\LX$, $P_\mathrm{spin}$, $PF$) combination shown by \belz\ is very similar to that of the magnetar 1E\,1048.1-5937. \cite{Tiengo:2005} reported for this source $P_\mathrm{spin}\simeq6.46$\,s, $PF\simeq91\%$ at a luminosity level of $\LX\simeq2.23\E{34}\ergs$ during \xmm\ obs. 0112780401 (Observation A in \citealt{Tiengo:2005}). Moreover, the spectral parameters they derived for the black bodies in the PL+BB and BB+BB models (first two models in Tab.~2 of the original article) are consistent with the ones we find for \belz\ in our spectral analysis. In particular, in the BB+BB model for the hot (cold) component they derive $R_\mathrm{BB}\simeq0.3$\,km (1.7\,km) and $kT_\mathrm{BB}\simeq1$\,keV (0.5\,keV, see Observation A in Tab.~2 of \citealt{Tiengo:2005}).

\section{Conclusions}\label{sec:conclusions}
%{\color{orange} Final remarks/conclusions.... including projects for the future [also depends on the outcomes of the optical spectra....\swift\ monitoring starting soon ]
%} 
We have reported on the discovery of a 7.25-s pulsar in the outskirts of the LMC, \belzlong\ (\belz)%, with a spin period $P_\mathrm{spin}=7.25$\,s
. %We have analysed the only available observation in the X-ray band in which the source has been detected, despite the fact that the field has been covered five more times by \xmm\ and \erosita\ in the last 5 years. 
%At the moment a \swift\ monitoring programme is active, to trigger new observations in case the source becomes active again.
%The position of the source in the LMC is supported by the intrinsic absorption ($N_\mathrm{H}\simeq4.8\E{21}\nH$) we derived from the fit of the X-ray spectrum in the 0.3--10\,keV band. 
%The emission is rather soft, peaking at $\sim1-2\,\mathrm{keV}$.
Based on \salt\ optical observations, we conclude that the only stellar object present within the X-ray uncertainty region is a faint%(relative magnitude $V\sim18.9\magnitude$)
, late-type (G8-K3 III) star% consistent with the distance of the LMC
. We discussed two scenarios for \belz: an accreting NS in a binary system with a late-type star companion (the first in the MCs) or an isolated NS, likely of the magnetar class. 
 
%Based on the properties of the known accreting X-ray pulsars in the MCs (all hosted in HMXBs) and 
%Under the hypothesis that the observed variability of \belz\ is related to the switching between the accretor and the propeller phase of the pulsar, we infer a dipolar magnetic field of $4.5\E{11}\,\mathrm{G}< B <9.1\E{11}\,\mathrm{G}$, in agreement with the values inferred for the bulk of the MCs population of accreting X-ray pulsars (all hosted in HMXBs). 
When comparing the \belz\ properties with those BeXRBs in the MCs and a few known Galactic X-ray pulsars in LMXBs/SyXBs with a similarly high magnetic field, the 0.3--10\,keV luminosity $\LX\simeq2.7\E{34}\ergs$ of \belz\ is too low for an accreting NS spinning at $\sim$7.25\,s and with the magnetic field strength reported above. Moreover, the pulsed fraction of the signal $PF\simeq86\%$ (with pulse minima consistent with null X-ray flux) is rarely observed in any accreting NS considered in our sample (see Fig.~\ref{fig:comparisonhist}). All these findings together make the accretion scenario unlikely. Nonetheless, if the accreting nature of \belz\ were confirmed, it would represent the outcome of a novel evolutionary path in the MCs. 

%The HMXB-scenario is also penalized by the fact that the only stellar object present within the X-ray uncertainty region is a faint (relative magnitude $V\sim18.9\magnitude$), late-type (G8-K3 III) star. It seems implausible, at this point, that \belz\ is a new exponent of the BeXRB class.\footnotetext{\url{http://astro.uni-tuebingen.de/~xrbcat/}.} At the same time, though unlikely, we cannot exclude the possibility that \belz\ is an accreting pulsar in a LMXB.  

On the other hand, the characteristics of \belz\ ($\LX$, $PF$ and $P_\mathrm{spin}$) are more commonly found in magnetars. In particular, another transient magnetar, 1E\,1048.1-5937, shows a very similar spin period $P_\mathrm{spin}\simeq6.46\,\mathrm{s}$ and similar pulsed fraction ($\simeq91\%$) at low luminosities $\LX\sim2\E{34}\ergs$ \citep{Tiengo:2005}. %It is also worth noticing that when a black body component is added to the spectral model, the derived parameters ($R_\mathrm{BB}\simeq0.6\,\mathrm{km}$, $kT_\mathrm{BB}\simeq0.6\,\mathrm{keV}$) are consistent with those usually found in magnetars during an outburst \citep[see][and references therein]{Esposito:2011,CotiMOOC:2018}. 
If the new pulsar were a magnetar, it would represent the third known magnetar outside our galaxy and the second in the LMC% after SGR\,0526--66 (one of the three objects for which a Giant Flare has been detected, see \citealt{Mazets:1979})
. 

An unambiguous classification of the new pulsar is complicated by the lack of a second measurement of the spin period which would have allowed us to infer the first derivative $\dot{P}_\mathrm{spin}$ of the spin period%, a parameter used to discriminate between the isolated (spin-down) and the accreting (spin-up) NS scenarios
. 
Although our findings point to a new member of the magnetar class, we cannot totally exclude, at this stage, the possibility that the source is a peculiar accreting NS in a LMXB/SyXB.
%Although the LMC is known to host at least another LMXB, i.e. LMC X-2 \citep{Agrawal:2009, Agrawal:2020}, \belz\ would represent the outcome of a never-seen-before evolutionary path in the MCs. Contrary to LMC X-2, for instance, the companion of \belz\ would be a red star, not a hot one. %\belz\ would also represent the fastest known NS in a LMXB with a G-K-type companion.
Future observations are needed in order to classify \belz. 
%which would represent an uncommon exponent of both classes we currently favour. 
%A classification of this source could also help to better understand the evolutionary path followed by the LMC that led to the observed abundance of X-ray sources. 

% The last numbered section should briefly summarise what has been done, and describe
% the final conclusions which the authors draw from their work.

\section*{Acknowledgements}
This work has %been partially supported by the ASI-INAF program I/004/11/5 and has 
made use of the XRT Data Analysis Software (XRTDAS) developed under the responsibility of the ASI Science Data Center (ASDC), Italy.
The optical observations reported in this paper were obtained with the Southern African Large Telescope (\salt). The \salt\ observations were obtained under the \salt\
Science Programme on the characterisation of the optical counterpart of the newly discovered pulsar (2022-2-SCI-012; PI: M.I.). This work has made use of data from the European Space Agency (ESA) mission
{\it Gaia} (\url{https://www.cosmos.esa.int/gaia}), processed by the {\it Gaia}
Data Processing and Analysis Consortium (DPAC,
\url{https://www.cosmos.esa.int/web/gaia/dpac/consortium}). Funding for the DPAC
has been provided by national institutions, in particular the institutions
participating in the {\it Gaia} Multilateral Agreement. This work is based on data from eROSITA, the soft X-ray instrument aboard SRG, a joint Russian-German science mission supported by the Russian Space Agency (Roskosmos), in the interests of the Russian Academy of Sciences represented by its Space Research Institute (IKI), and the Deutsches Zentrum f\"ur Luft- und Raumfahrt (DLR). The SRG spacecraft was built by Lavochkin Association (NPOL) and its subcontractors, and is operated by NPOL with support from the Max Planck Institute for Extraterrestrial Physics (MPE). The development and construction of the eROSITA X-ray instrument was led by MPE, with contributions from the Dr. Karl Remeis Observatory Bamberg \& ECAP (FAU Erlangen-Nuernberg), the University of Hamburg Observatory, the Leibniz Institute for Astrophysics Potsdam (AIP), and the Institute for Astronomy and Astrophysics of the University of T\"ubingen, with the support of DLR and the Max Planck Society. The Argelander Institute for Astronomy of the University of Bonn and the Ludwig Maximilians Universit\"at Munich also participated in the science preparation for eROSITA. \textsc{iraf} was distributed by the National Optical Astronomy Observatory, which was managed by the Association of Universities for Research in Astronomy (AURA) under a cooperative agreement with the National Science Foundation.
M.I. is supported by the AASS Ph.D. joint research programme between the University of Rome "Sapienza" and the University of Rome "Tor Vergata", with the collaboration of the National Institute of Astrophysics (INAF). M.I. thanks Dr. Elizabeth Naluminsa for the on-site technical support with the \salt\ observations. G.L.I., P.E., and A.T. acknowledge financial support from the Italian Ministry for University and Research, through grant 2017LJ39LM (UNIAM). F.C.Z. is supported by a Ram\'on y Cajal fellowship (grant agreement RYC2021-030888-I). F.C.Z. and N.R. are supported by the ERC Consolidator Grant "MAGNESIA" (No. 817661) and are also partially supported by the program Unidad de Excelencia Mar\'ia de Maeztu CEX2020-001058-M. L.S. acknowledges financial contributions from ASI-INAF agreements 2017-14-H.O and  I/037/12/0; from “iPeska” research grant (P.I. Andrea Possenti) funded under the INAF call PRIN-SKA/CTA (resolution 70/2016), 
 from PRIN-INAF 2019 no. 15 and from the Italian Ministry of University and Research (MUR), PRIN 2020 (prot. 2020BRP57Z) ‘Gravitational and Electromagnetic-wave Sources in the Universe with current and nextgeneration detectors (GEMS)’.

\emph{Software}: Matplotlib \citep{Matplotlib:2007}, numpy \citep{Numpy:2020}, pandas \citep{McKinney:2010}, \textsc{sas} \citep{SAS:2004}, \textsc{heasoft}, EXTraS pipeline \citep{DeLuca:2021}, \textsc{iraf}.%{\color{blue} Ringraziare il team di \erosita.}
% The Acknowledgements section is not numbered. Here you can thank helpful
% colleagues, acknowledge funding agencies, telescopes and facilities used etc.
% Try to keep it short.

%%%%%%%%%%%%%%%%%%%%%%%%%%%%%%%%%%%%%%%%%%%%%%%%%%
\section*{Data Availability}

% The inclusion of a Data Availability Statement is a requirement for articles published in MNRAS. Data Availability Statements provide a standardised format for readers to understand the availability of data underlying the research results described in the article. The statement may refer to original data generated in the course of the study or to third-party data analysed in the article. The statement should describe and provide means of access, where possible, by linking to the data or providing the required accession numbers for the relevant databases or DOIs.

The \xmm\ data analysed in this article are public and can be downloaded from the \xmm\ Science Archive XSA (\texttt{http://nxsa.esac.esa.int/nxsa-web/\#search}) and the High Energy Astrophysics Science Archive Research Center (HEASARC) archive (\texttt{https://heasarc.gsfc.nasa.gov/cgi\-bin/W3Browse/w3browse.pl}). \salt\ data are available upon request to the authors. 

%%%%%%%%%%%%%%%%%%%% REFERENCES %%%%%%%%%%%%%%%%%%

% The best way to enter references is to use BibTeX:

\bibliographystyle{mnras}
%\bibliography{example} % if your bibtex file is called example.bib
\bibliography{references_ads}

\begin{thebibliography}{}
\makeatletter
\relax
\def\mn@urlcharsother{\let\do\@makeother \do\$\do\&\do\#\do\^\do\_\do\%\do\~}
\def\mn@doi{\begingroup\mn@urlcharsother \@ifnextchar [ {\mn@doi@}
  {\mn@doi@[]}}
\def\mn@doi@[#1]#2{\def\@tempa{#1}\ifx\@tempa\@empty \href
  {http://dx.doi.org/#2} {doi:#2}\else \href {http://dx.doi.org/#2} {#1}\fi
  \endgroup}
\def\mn@eprint#1#2{\mn@eprint@#1:#2::\@nil}
\def\mn@eprint@arXiv#1{\href {http://arxiv.org/abs/#1} {{\tt arXiv:#1}}}
\def\mn@eprint@dblp#1{\href {http://dblp.uni-trier.de/rec/bibtex/#1.xml}
  {dblp:#1}}
\def\mn@eprint@#1:#2:#3:#4\@nil{\def\@tempa {#1}\def\@tempb {#2}\def\@tempc
  {#3}\ifx \@tempc \@empty \let \@tempc \@tempb \let \@tempb \@tempa \fi \ifx
  \@tempb \@empty \def\@tempb {arXiv}\fi \@ifundefined
  {mn@eprint@\@tempb}{\@tempb:\@tempc}{\expandafter \expandafter \csname
  mn@eprint@\@tempb\endcsname \expandafter{\@tempc}}}

\bibitem[\protect\citeauthoryear{{Agrawal} \& {Misra}}{{Agrawal} \&
  {Misra}}{2009}]{Agrawal:2009}
{Agrawal} V.~K.,  {Misra} R.,  2009, \mn@doi [MNRAS]
  {10.1111/j.1365-2966.2009.15014.x}, \href
  {https://ui.adsabs.harvard.edu/abs/2009MNRAS.398.1352A} {398, 1352}

\bibitem[\protect\citeauthoryear{{Agrawal} \& {Nandi}}{{Agrawal} \&
  {Nandi}}{2020}]{Agrawal:2020}
{Agrawal} V.~K.,  {Nandi} A.,  2020, \mn@doi [MNRAS] {10.1093/mnras/staa2063},
  \href {https://ui.adsabs.harvard.edu/abs/2020MNRAS.497.3726A} {497, 3726}

\bibitem[\protect\citeauthoryear{{{\'A}lvarez-Hern{\'a}ndez}
  et~al.,}{{{\'A}lvarez-Hern{\'a}ndez} et~al.}{2021}]{GKPersPorb:2021}
{{\'A}lvarez-Hern{\'a}ndez} A.,  et~al., 2021, \mn@doi [MNRAS]
  {10.1093/mnras/stab2547}, \href
  {https://ui.adsabs.harvard.edu/abs/2021MNRAS.507.5805A} {507, 5805}

\bibitem[\protect\citeauthoryear{Antoniou \& Zezas}{Antoniou \&
  Zezas}{2016}]{Antoniou:2016}
Antoniou V.,  Zezas A.,  2016, \mn@doi [MNRAS] {10.1093/MNRAS/STW167}, \href
  {https://ui.adsabs.harvard.edu/abs/2016MNRAS.459..528A} {459, 528}

\bibitem[\protect\citeauthoryear{Antoniou, Zezas, Hatzidimitriou  \&
  Kalogera}{Antoniou et~al.}{2010}]{Antoniou:2010}
Antoniou V.,  Zezas A.,  Hatzidimitriou D.,   Kalogera V.,  2010, \mn@doi [ApJ]
  {10.1088/2041-8205/716/2/L140}, \href
  {https://ui.adsabs.harvard.edu/abs/2010ApJ...716L.140A} {716, L140}

\bibitem[\protect\citeauthoryear{Arnaud}{Arnaud}{1996}]{Arnaud:1996}
Arnaud K.~A.,  1996, in Jacoby G.~H.,  Barnes J.,  eds, ~ASPC Vol. 101,
  Astronomical Data Analysis Software and Systems V. p.~17

\bibitem[\protect\citeauthoryear{{Avakyan}, {Neumann}, {Zainab}, {Doroshenko},
  {Wilms}  \& {Santangelo}}{{Avakyan} et~al.}{2023}]{XRBCat:LMXB}
{Avakyan} A.,  {Neumann} M.,  {Zainab} A.,  {Doroshenko} V.,  {Wilms} J.,
  {Santangelo} A.,  2023, \mn@doi [arXiv e-prints] {10.48550/arXiv.2303.16168},
  \href {https://ui.adsabs.harvard.edu/abs/2023arXiv230316168A} {p.
  arXiv:2303.16168}

\bibitem[\protect\citeauthoryear{{Bahramian} \& {Degenaar}}{{Bahramian} \&
  {Degenaar}}{2022}]{Bahramian:2022}
{Bahramian} A.,  {Degenaar} N.,  2022, \mn@doi [arXiv e-prints]
  {10.48550/arXiv.2206.10053}, \href
  {https://ui.adsabs.harvard.edu/abs/2022arXiv220610053B} {p. arXiv:2206.10053}

\bibitem[\protect\citeauthoryear{Bartlett, Coe, Israel, Clark, Esposito, D'Elia
   \& Udalski}{Bartlett et~al.}{2017}]{Bartlett:2017}
Bartlett E.~S.,  Coe M.~J.,  Israel G.~L.,  Clark J.~S.,  Esposito P.,  D'Elia
  V.,   Udalski A.,  2017, \mn@doi [MNRAS] {10.1093/MNRAS/STX032}, \href
  {https://ui.adsabs.harvard.edu/abs/2017MNRAS.466.4659B} {466, 4659}

\bibitem[\protect\citeauthoryear{Boon et~al.,}{Boon et~al.}{2017}]{Boon:2017}
Boon C.~M.,  et~al., 2017, \mn@doi [MNRAS] {10.1093/mnras/stw3169}, \href
  {https://ui.adsabs.harvard.edu/abs/2017MNRAS.466.1149B} {466, 1149}

\bibitem[\protect\citeauthoryear{Buckley, Swart  \& Meiring}{Buckley
  et~al.}{2006}]{Buckley:2006}
Buckley D. A.~H.,  Swart G.~P.,   Meiring J.~G.,  2006, in Stepp L.~M.,  ed.,
  SPIE. p. 62670Z, \mn@doi{10.1117/12.673750}

\bibitem[\protect\citeauthoryear{Burgh, Nordsieck, Kobulnicky, Williams,
  O'Donoghue, Smith  \& Percival}{Burgh et~al.}{2003}]{Burgh:2003}
Burgh E.~B.,  Nordsieck K.~H.,  Kobulnicky H.~A.,  Williams T.~B.,  O'Donoghue
  D.,  Smith M.~P.,   Percival J.~W.,  2003, in Iye M.,  Moorwood A. F.~M.,
  eds, SPIE. p.~1463, \mn@doi{10.1117/12.460312}

\bibitem[\protect\citeauthoryear{Carpano, Haberl, Maitra, Freyberg, Dennerl,
  Schwope, Buckley  \& Monageng}{Carpano et~al.}{2022}]{Carpano:2022}
Carpano S.,  Haberl F.,  Maitra C.,  Freyberg M.,  Dennerl K.,  Schwope A.,
  Buckley A.~H.,   Monageng I.~M.,  2022, \mn@doi [A{\&}A]
  {10.1051/0004-6361/202141082}, \href
  {https://ui.adsabs.harvard.edu/abs/2022A&A...661A..20C} {661, A20}

\bibitem[\protect\citeauthoryear{Chatterjee, Agrawal  \& Nandi}{Chatterjee
  et~al.}{2021}]{Chatterjee:2021}
Chatterjee R.,  Agrawal V.~K.,   Nandi A.,  2021, \mn@doi [MNRAS]
  {10.1093/mnras/stab1499}, \href
  {https://ui.adsabs.harvard.edu/abs/2021MNRAS.505.3785C} {505, 3785}

\bibitem[\protect\citeauthoryear{Christodoulou, Laycock, Yang  \&
  Fingerman}{Christodoulou et~al.}{2016}]{Christodoulou:2016}
Christodoulou D.~M.,  Laycock S. G.~T.,  Yang J.,   Fingerman S.,  2016,
  \mn@doi [ApJ] {10.3847/0004-637X/829/1/30}, \href
  {https://ui.adsabs.harvard.edu/abs/2016ApJ...829...30C} {829, 30}

\bibitem[\protect\citeauthoryear{Coe}{Coe}{2000}]{Coe:2000}
Coe M.~J.,  2000, \mn@doi [ASPC] {10.48550/arXiv.astro-ph/9911272}, \href
  {https://ui.adsabs.harvard.edu/abs/2000ASPC..214..656C} {214, 656}

\bibitem[\protect\citeauthoryear{Coe \& Kirk}{Coe \& Kirk}{2015}]{Coe:2015}
Coe M.~J.,  Kirk J.,  2015, \mn@doi [MNRAS] {10.1093/MNRAS/STV1283}, \href
  {https://ui.adsabs.harvard.edu/abs/2015MNRAS.452..969C} {452, 969}

\bibitem[\protect\citeauthoryear{Coti~Zelati, Rea, Pons, Campana  \&
  Esposito}{Coti~Zelati et~al.}{2018}]{CotiMOOC:2018}
Coti~Zelati F.,  Rea N.,  Pons J.~A.,  Campana S.,   Esposito P.,  2018,
  \mn@doi [MNRAS] {10.1093/MNRAS/STX2679}, \href
  {https://ui.adsabs.harvard.edu/abs/2018MNRAS.474..961C} {474, 961}

\bibitem[\protect\citeauthoryear{Crawford et~al.,}{Crawford
  et~al.}{2012}]{Crawford:2012}
Crawford S.~M.,  et~al., 2012, {PySALT: SALT science pipeline}, Astrophysics
  Source Code Library, record ascl:1207.010

\bibitem[\protect\citeauthoryear{De~Luca et~al.,}{De~Luca
  et~al.}{2021}]{DeLuca:2021}
De~Luca A.,  et~al., 2021, \mn@doi [A{\&}A] {10.1051/0004-6361/202039783},
  \href {https://ui.adsabs.harvard.edu/abs/2021A&A...650A.167D} {650, 167}

\bibitem[\protect\citeauthoryear{Dickey \& Lockman}{Dickey \&
  Lockman}{1990}]{DickeyHI:1990}
Dickey J.~M.,  Lockman F.~J.,  1990, \mn@doi [ARA{\&}A]
  {10.1146/annurev.aa.28.090190.001243}, \href
  {https://ui.adsabs.harvard.edu/abs/1990ARA&A..28..215D} {28, 215}

\bibitem[\protect\citeauthoryear{Duncan \& Thompson}{Duncan \&
  Thompson}{1992}]{Duncan:1992}
Duncan R.~C.,  Thompson C.,  1992, \mn@doi [ApJ] {10.1086/186413}, \href
  {https://ui.adsabs.harvard.edu/abs/1992ApJ...392L...9D} {392, L9}

\bibitem[\protect\citeauthoryear{{Esposito}, {Rea}  \& {Israel}}{{Esposito}
  et~al.}{2021}]{Esposito:2021}
{Esposito} P.,  {Rea} N.,   {Israel} G.~L.,  2021, in {Belloni} T.~M.,
  {M{\'e}ndez} M.,   {Zhang} C.,  eds,  Astrophysics and Space Science Library
  Vol. 461, Timing Neutron Stars: Pulsations, Oscillations and Explosions. pp
  97--142 (\mn@eprint {arXiv} {1803.05716}),
  \mn@doi{10.1007/978-3-662-62110-3_3}

\bibitem[\protect\citeauthoryear{Gabriel et~al.,}{Gabriel
  et~al.}{2004}]{SAS:2004}
Gabriel C.,  et~al., 2004, in Ochsenbein F.,  Allen M.~G.,   Egret D.,  eds,
  ~ASPC Vol. 314, Astronomical Data Analysis Software and Systems (ADASS) XIII.
  p.~759

\bibitem[\protect\citeauthoryear{{Gaia Collaboration} et~al.,}{{Gaia
  Collaboration} et~al.}{2022}]{GaiaDR3:2022}
{Gaia Collaboration} et~al., 2022, \mn@doi [arXiv e-prints]
  {10.48550/arXiv.2208.00211}, \href
  {https://ui.adsabs.harvard.edu/abs/2022arXiv220800211G} {p. arXiv:2208.00211}

\bibitem[\protect\citeauthoryear{Gotthelf, Halpern, Mori  \&
  Beloborodov}{Gotthelf et~al.}{2019}]{Gotthelf:2019}
Gotthelf E.~V.,  Halpern J.~P.,  Mori K.,   Beloborodov A.~M.,  2019, \mn@doi
  [ApJ] {10.3847/1538-4357/AB378C}, \href
  {https://ui.adsabs.harvard.edu/abs/2019ApJ...882..173G} {882, 173}

\bibitem[\protect\citeauthoryear{Graczyk et~al.,}{Graczyk
  et~al.}{2014}]{Graczyk:2014}
Graczyk D.,  et~al., 2014, \mn@doi [ApJ] {10.1088/0004-637X/780/1/59}, \href
  {https://ui.adsabs.harvard.edu/abs/2014ApJ...780...59G} {780, 59}

\bibitem[\protect\citeauthoryear{Haberl \& Sturm}{Haberl \&
  Sturm}{2016}]{Haberl:2016}
Haberl F.,  Sturm R.,  2016, \mn@doi [A{\&}A] {10.1051/0004-6361/201527326},
  \href {https://ui.adsabs.harvard.edu/abs/2016A&A...586A..81H} {586, A81}

\bibitem[\protect\citeauthoryear{Haberl et~al.,}{Haberl
  et~al.}{2017}]{HaberlF:2017}
Haberl F.,  et~al., 2017, \mn@doi [A{\&}A] {10.1051/0004-6361/201629744}, \href
  {https://ui.adsabs.harvard.edu/abs/2017A&A...598A..69H} {598, A69}

\bibitem[\protect\citeauthoryear{Haberl et~al.,}{Haberl
  et~al.}{2022a}]{HaberlSXP304:2022}
Haberl F.,  et~al., 2022a, \mn@doi [A{\&}A] {10.1051/0004-6361/202141878},
  \href {https://ui.adsabs.harvard.edu/abs/2022A&A...661A..25H} {661, A25}

\bibitem[\protect\citeauthoryear{Haberl, Maitra, Vasilopoulos, Maggi, Udalski,
  Monageng  \& Buckley}{Haberl et~al.}{2022b}]{Haberl:2022}
Haberl F.,  Maitra C.,  Vasilopoulos G.,  Maggi P.,  Udalski A.,  Monageng
  I.~M.,   Buckley D. A.~H.,  2022b, \mn@doi [A{\&}A]
  {10.1051/0004-6361/202243301}, \href
  {https://ui.adsabs.harvard.edu/abs/2022A&A...662A..22H} {662, A22}

\bibitem[\protect\citeauthoryear{{Haberl} et~al.,}{{Haberl}
  et~al.}{2023a}]{HaberlLMXB:2023}
{Haberl} F.,  et~al., 2023a, \mn@doi [A\&A] {10.1051/0004-6361/202245015},
  \href {https://ui.adsabs.harvard.edu/abs/2023A&A...669A..66H} {669, A66}

\bibitem[\protect\citeauthoryear{Haberl et~al.,}{Haberl
  et~al.}{2023b}]{Haberl:2023}
Haberl F.,  et~al., 2023b, \mn@doi [A{\&}A] {10.1051/0004-6361/202245807},
  \href {https://ui.adsabs.harvard.edu/abs/2023A&A...671A..90H} {671, A90}

\bibitem[\protect\citeauthoryear{{Harris} \& {Zaritsky}}{{Harris} \&
  {Zaritsky}}{2009}]{Harris:2009}
{Harris} J.,  {Zaritsky} D.,  2009, \mn@doi [AJ]
  {10.1088/0004-6256/138/5/1243}, \href
  {https://ui.adsabs.harvard.edu/abs/2009AJ....138.1243H} {138, 1243}

\bibitem[\protect\citeauthoryear{Harris et~al.,}{Harris
  et~al.}{2020}]{Numpy:2020}
Harris C.~R.,  et~al., 2020, \mn@doi [Nature] {10.1038/s41586-020-2649-2},
  \href {https://ui.adsabs.harvard.edu/abs/2020Natur.585..357H} {585, 357}

\bibitem[\protect\citeauthoryear{Hunter}{Hunter}{2007}]{Matplotlib:2007}
Hunter J.~D.,  2007, \mn@doi [CSE] {10.1109/MCSE.2007.55}, \href
  {https://ui.adsabs.harvard.edu/abs/2007CSE.....9...90H} {9, 90}

\bibitem[\protect\citeauthoryear{Israel \& Stella}{Israel \&
  Stella}{1996}]{Israel:1996}
Israel G.~L.,  Stella L.,  1996, \mn@doi [ApJ] {10.1086/177697}, \href
  {https://ui.adsabs.harvard.edu/abs/1996ApJ...468..369I} {468, 369}

\bibitem[\protect\citeauthoryear{Israel, Stella, Angelini, White, Kallman,
  Giommi  \& Treves}{Israel et~al.}{1997}]{Israel:1997}
Israel G.~L.,  Stella L.,  Angelini L.,  White N.~E.,  Kallman T.~R.,  Giommi
  P.,   Treves A.,  1997, \mn@doi [ApJ] {10.1086/310418}, \href
  {https://ui.adsabs.harvard.edu/abs/1997ApJ...474L..53I} {474, L53}

\bibitem[\protect\citeauthoryear{Israel et~al.,}{Israel
  et~al.}{2016}]{IsraelSGR1935:2016}
Israel G.~L.,  et~al., 2016, \mn@doi [MNRAS] {10.1093/MNRAS/STW008}, \href
  {https://ui.adsabs.harvard.edu/abs/2016MNRAS.457.3448I} {457, 3448}

\bibitem[\protect\citeauthoryear{{Jahoda}, {Markwardt}, {Radeva}, {Rots},
  {Stark}, {Swank}, {Strohmayer}  \& {Zhang}}{{Jahoda}
  et~al.}{2006}]{Jahoda:2006}
{Jahoda} K.,  {Markwardt} C.~B.,  {Radeva} Y.,  {Rots} A.~H.,  {Stark} M.~J.,
  {Swank} J.~H.,  {Strohmayer} T.~E.,   {Zhang} W.,  2006, \mn@doi [ApJS]
  {10.1086/500659}, \href
  {https://ui.adsabs.harvard.edu/abs/2006ApJS..163..401J} {163, 401}

\bibitem[\protect\citeauthoryear{Jansen et~al.,}{Jansen
  et~al.}{2001}]{Jansen:2001}
Jansen F.,  et~al., 2001, \mn@doi [A{\&}A] {10.1051/0004-6361:20000036}, \href
  {https://ui.adsabs.harvard.edu/abs/2001A&A...365L...1J} {365, L1}

\bibitem[\protect\citeauthoryear{Kaspi \& Beloborodov}{Kaspi \&
  Beloborodov}{2017}]{Kaspi:2017}
Kaspi V.~M.,  Beloborodov A.~M.,  2017, \mn@doi [ARA{\&}A]
  {10.1146/annurev-astro-081915-023329}, \href
  {https://ui.adsabs.harvard.edu/abs/2017ARA&A..55..261K} {55, 261}

\bibitem[\protect\citeauthoryear{Kennea, Evans, Coe  \& Udalski}{Kennea
  et~al.}{2017}]{Kennea:2017}
Kennea J.~A.,  Evans P.~A.,  Coe M.~J.,   Udalski A.,  2017, ATEL, \href
  {https://ui.adsabs.harvard.edu/abs/2017ATel10600....1K} {10600, 1}

\bibitem[\protect\citeauthoryear{Kirsch et~al.,}{Kirsch
  et~al.}{2004}]{XMMcalibration:2004}
Kirsch M.,  et~al., 2004, \mn@doi [SPIE] {10.1117/12.549276}, \href
  {https://ui.adsabs.harvard.edu/abs/2004SPIE.5488..103K} {5488, 103}

\bibitem[\protect\citeauthoryear{Kobulnicky, Nordsieck, Burgh, Smith, Percival,
  Williams  \& O'Donoghue}{Kobulnicky et~al.}{2003}]{Kobulnicky:2003}
Kobulnicky H.~A.,  Nordsieck K.~H.,  Burgh E.~B.,  Smith M.~P.,  Percival
  J.~W.,  Williams T.~B.,   O'Donoghue D.,  2003, in Iye M.,  Moorwood A.
  F.~M.,  eds, SPIE. p.~1634, \mn@doi{10.1117/12.460315}

\bibitem[\protect\citeauthoryear{Koliopanos \& Vasilopoulos}{Koliopanos \&
  Vasilopoulos}{2018}]{Koliopanos:2018}
Koliopanos F.,  Vasilopoulos G.,  2018, \mn@doi [A{\&}A]
  {10.1051/0004-6361/201731623}, \href
  {https://ui.adsabs.harvard.edu/abs/2018A&A...614A..23K} {614, A23}

\bibitem[\protect\citeauthoryear{Kulkarni, Kaplan, Marshall, Frail, Murakami
  \& Yonetoku}{Kulkarni et~al.}{2003}]{Kulkarni:2003}
Kulkarni S.~R.,  Kaplan D.~L.,  Marshall H.~L.,  Frail D.~A.,  Murakami T.,
  Yonetoku D.,  2003, \mn@doi [ApJ] {10.1086/346110}, \href
  {https://ui.adsabs.harvard.edu/abs/2003ApJ...585..948K} {585, 948}

\bibitem[\protect\citeauthoryear{La~Palombara, Esposito, Mereghetti, Pintore,
  Sidoli  \& Tiengo}{La~Palombara et~al.}{2018a}]{LaPalombara:2018}
La~Palombara N.,  Esposito P.,  Mereghetti S.,  Pintore F.,  Sidoli L.,
  Tiengo A.,  2018a, \mn@doi [MNRAS] {10.1093/MNRAS/STX3283}, \href
  {https://ui.adsabs.harvard.edu/abs/2018MNRAS.475.1382L} {475, 1382}

\bibitem[\protect\citeauthoryear{La~Palombara, Esposito, Pintore, Sidoli,
  Mereghetti  \& Tiengo}{La~Palombara et~al.}{2018b}]{LaPalombaraSXP59:2018}
La~Palombara N.,  Esposito P.,  Pintore F.,  Sidoli L.,  Mereghetti S.,
  Tiengo A.,  2018b, \mn@doi [A{\&}A] {10.1051/0004-6361/201833907}, \href
  {https://ui.adsabs.harvard.edu/abs/2018A&A...619A.126L} {619, A126}

\bibitem[\protect\citeauthoryear{Lamb, Fox, Macomb  \& Prince}{Lamb
  et~al.}{2002}]{Lamb:2002}
Lamb R.~C.,  Fox D.~W.,  Macomb D.~J.,   Prince T.~A.,  2002, \mn@doi [ApJ]
  {10.1086/342352}, \href
  {https://ui.adsabs.harvard.edu/abs/2002ApJ...574L..29L} {574, L29}

\bibitem[\protect\citeauthoryear{Lamb, Fox, Macomb  \& Prince}{Lamb
  et~al.}{2003}]{LambAddendum:2003}
Lamb R.~C.,  Fox D.~W.,  Macomb D.~J.,   Prince T.~A.,  2003, \mn@doi [ApJ]
  {10.1086/381175}, \href
  {https://ui.adsabs.harvard.edu/abs/2003ApJ...599L.115L} {599, L115}

\bibitem[\protect\citeauthoryear{{Lavagetto}, {Iaria}, {D'A{\i}}, {di Salvo}
  \& {Robba}}{{Lavagetto} et~al.}{2008}]{Lavagetto:2008}
{Lavagetto} G.,  {Iaria} R.,  {D'A{\i}} A.,  {di Salvo} T.,   {Robba} N.~R.,
  2008, \mn@doi [A{\&}A] {10.1051/0004-6361:20078027}, \href
  {https://ui.adsabs.harvard.edu/abs/2008A&A...478..181L} {478, 181}

\bibitem[\protect\citeauthoryear{Laycock, Corbet, Perrodin, Coe, Marshall  \&
  Markwardt}{Laycock et~al.}{2002}]{Laycock:2002}
Laycock S.,  Corbet R. H.~D.,  Perrodin D.,  Coe M.~J.,  Marshall F.~E.,
  Markwardt C.,  2002, \mn@doi [A{\&}A] {10.1051/0004-6361:20020150}, \href
  {https://ui.adsabs.harvard.edu/abs/2002A&A...385..464L} {385, 464}

\bibitem[\protect\citeauthoryear{Leahy, Darbro, Elsner, Weisskopf, Sutherland,
  Kahn  \& Grindlay}{Leahy et~al.}{1983}]{Leahy:1983}
Leahy D.~A.,  Darbro W.,  Elsner R.~F.,  Weisskopf M.~C.,  Sutherland P.~G.,
  Kahn S.,   Grindlay J.~E.,  1983, \mn@doi [ApJ] {10.1086/161288}, \href
  {https://ui.adsabs.harvard.edu/abs/1983ApJ...272..256L} {266, 160}

\bibitem[\protect\citeauthoryear{{Li}, {Torres}, {Rea}, {de O{\~n}a Wilhelmi},
  {Papitto}, {Hou}  \& {Mauche}}{{Li} et~al.}{2016}]{LiAEAqr:2016}
{Li} J.,  {Torres} D.~F.,  {Rea} N.,  {de O{\~n}a Wilhelmi} E.,  {Papitto} A.,
  {Hou} X.,   {Mauche} C.~W.,  2016, \mn@doi [ApJ]
  {10.3847/0004-637X/832/1/35}, \href
  {https://ui.adsabs.harvard.edu/abs/2016ApJ...832...35L} {832, 35}

\bibitem[\protect\citeauthoryear{Maitra, Paul, Haberl  \& Vasilopoulos}{Maitra
  et~al.}{2018}]{Maitra:2018}
Maitra C.,  Paul B.,  Haberl F.,   Vasilopoulos G.,  2018, \mn@doi [MNRAS]
  {10.1093/mnrasl/sly141}, \href
  {https://ui.adsabs.harvard.edu/abs/2018MNRAS.480L.136M} {480, L136}

\bibitem[\protect\citeauthoryear{Maitra, Haberl, Maggi, Kavanagh, Vasilopoulos,
  Sasaki, Filipovi{\'{c}}  \& Udalski}{Maitra et~al.}{2021}]{Maitra:2021}
Maitra C.,  Haberl F.,  Maggi P.,  Kavanagh P.~J.,  Vasilopoulos G.,  Sasaki
  M.,  Filipovi{\'{c}} M.~D.,   Udalski A.,  2021, \mn@doi [MNRAS]
  {10.1093/MNRAS/STAB716}, \href
  {https://ui.adsabs.harvard.edu/abs/2021MNRAS.504..326M} {504, 326}

\bibitem[\protect\citeauthoryear{Maitra et~al.,}{Maitra
  et~al.}{2023}]{Maitra:2022}
Maitra C.,  et~al., 2023, \mn@doi [A{\&}A] {10.1051/0004-6361/202244328}, \href
  {https://ui.adsabs.harvard.edu/abs/2023A&A...669A..30M} {669, A30}

\bibitem[\protect\citeauthoryear{Mazets, Golenetskii, Il'inskii, Aptekar'  \&
  Guryan}{Mazets et~al.}{1979}]{Mazets:1979}
Mazets E.~P.,  Golenetskii S.~V.,  Il'inskii V.~N.,  Aptekar' R.~L.,   Guryan
  Y.~A.,  1979, \mn@doi [Nature] {10.1038/282587a0}, \href
  {https://ui.adsabs.harvard.edu/abs/1979Natur.282..587M} {282, 587}

\bibitem[\protect\citeauthoryear{McGarry, Gaensler, Ransom, Kaspi  \&
  Veljkovik}{McGarry et~al.}{2005}]{McGarry:2005}
McGarry M.~B.,  Gaensler B.~M.,  Ransom S.~M.,  Kaspi V.~M.,   Veljkovik S.,
  2005, \mn@doi [ApJ] {10.1086/432441}, \href
  {https://ui.adsabs.harvard.edu/abs/2005ApJ...627L.137M} {627, L137}

\bibitem[\protect\citeauthoryear{McKinney}{McKinney}{2010}]{McKinney:2010}
McKinney W.,  2010, in van~der Walt S.,  Millman J.,  eds, Proceedings of the
  9th Python in Science Conference. Proceedings of the Python in Science
  Conference.
SciPy, pp 56--61, \mn@doi{10.25080/Majora-92bf1922-00a}

\bibitem[\protect\citeauthoryear{Mereghetti et~al.,}{Mereghetti
  et~al.}{2006}]{Mereghetti:2006}
Mereghetti S.,  et~al., 2006, \mn@doi [ApJ] {10.1086/508682}, \href
  {https://ui.adsabs.harvard.edu/abs/2006ApJ...653.1423M} {653, 1423}

\bibitem[\protect\citeauthoryear{Mereghetti et~al.,}{Mereghetti
  et~al.}{2021}]{Mereghetti:2021}
Mereghetti S.,  et~al., 2021, \mn@doi [MNRAS] {10.1093/mnras/stab1004}, \href
  {https://ui.adsabs.harvard.edu/abs/2021MNRAS.504..920M} {504, 920}

\bibitem[\protect\citeauthoryear{Okazaki \& Negueruela}{Okazaki \&
  Negueruela}{2001}]{Okazaki:2001}
Okazaki A.~T.,  Negueruela I.,  2001, \mn@doi [A{\&}A]
  {10.1051/0004-6361:20011083}, \href
  {https://ui.adsabs.harvard.edu/abs/2001A&A...377..161O} {377, 161}

\bibitem[\protect\citeauthoryear{Olausen \& Kaspi}{Olausen \&
  Kaspi}{2014}]{McGill:catalogue}
Olausen S.~A.,  Kaspi V.~M.,  2014, \mn@doi [ApJS] {10.1088/0067-0049/212/1/6},
  \href {https://ui.adsabs.harvard.edu/abs/2014ApJS..212....6O} {212, 6}

\bibitem[\protect\citeauthoryear{Park, Bhalerao, Kargaltsev  \& Slane}{Park
  et~al.}{2020}]{Park:2020}
Park S.,  Bhalerao J.,  Kargaltsev O.,   Slane P.~O.,  2020, \mn@doi [ApJ]
  {10.3847/1538-4357/ab83f8}, \href
  {https://ui.adsabs.harvard.edu/abs/2020ApJ...894...17P} {894, 17}

\bibitem[\protect\citeauthoryear{Pelisoli et~al.,}{Pelisoli
  et~al.}{2022}]{Pelisoli:2022}
Pelisoli I.,  et~al., 2022, \mn@doi [MNRAS] {10.1093/mnrasl/slab116}, \href
  {https://ui.adsabs.harvard.edu/abs/2022MNRAS.509L..31P} {509, L31}

\bibitem[\protect\citeauthoryear{{Piatti}, {Hwang}, {Cole}, {Angelo}  \&
  {Emptage}}{{Piatti} et~al.}{2018}]{Piatti:2018}
{Piatti} A.~E.,  {Hwang} N.,  {Cole} A.~A.,  {Angelo} M.~S.,   {Emptage} B.,
  2018, \mn@doi [MNRAS] {10.1093/mnras/sty2324}, \href
  {https://ui.adsabs.harvard.edu/abs/2018MNRAS.481...49P} {481, 49}

\bibitem[\protect\citeauthoryear{Pietrzy{\'{n}}ski et~al.,}{Pietrzy{\'{n}}ski
  et~al.}{2013}]{LMCdistance:2013}
Pietrzy{\'{n}}ski G.,  et~al., 2013, \mn@doi [Nature] {10.1038/nature11878},
  \href {https://ui.adsabs.harvard.edu/abs/2013Natur.495...76P} {495, 76}

\bibitem[\protect\citeauthoryear{Predehl et~al.,}{Predehl
  et~al.}{2021}]{Predhel:2021}
Predehl P.,  et~al., 2021, \mn@doi [A{\&}A] {10.1051/0004-6361/202039313},
  \href {https://ui.adsabs.harvard.edu/abs/2021A&A...647A...1P} {647, A1}

\bibitem[\protect\citeauthoryear{Prusti et~al.,}{Prusti
  et~al.}{2016}]{GaiaMission:2016}
Prusti T.,  et~al., 2016, \mn@doi [A{\&}A] {10.1051/0004-6361/201629272}, \href
  {https://ui.adsabs.harvard.edu/abs/2016A&A...595A...1G} {595, A1}

\bibitem[\protect\citeauthoryear{Rea et~al.,}{Rea et~al.}{2007a}]{Rea:2007}
Rea N.,  et~al., 2007a, \mn@doi [Ap{\&}SS] {10.1007/S10509-007-9310-5}, \href
  {https://ui.adsabs.harvard.edu/abs/2007Ap&SS.308..505R} {308, 505}

\bibitem[\protect\citeauthoryear{Rea et~al.,}{Rea
  et~al.}{2007b}]{ReaMNRAS:2007}
Rea N.,  et~al., 2007b, \mn@doi [MNRAS] {10.1111/J.1365-2966.2007.12257.X},
  \href {https://ui.adsabs.harvard.edu/abs/2007MNRAS.381..293R} {381, 293}

\bibitem[\protect\citeauthoryear{{Saito}, {Baptista}, {Horne}  \&
  {Martell}}{{Saito} et~al.}{2010}]{Saito:2010}
{Saito} R.~K.,  {Baptista} R.,  {Horne} K.,   {Martell} P.,  2010, \mn@doi [AJ]
  {10.1088/0004-6256/139/6/2542}, \href
  {https://ui.adsabs.harvard.edu/abs/2010AJ....139.2542S} {139, 2542}

\bibitem[\protect\citeauthoryear{{Schlafly} \& {Finkbeiner}}{{Schlafly} \&
  {Finkbeiner}}{2011}]{Schlafly:2011}
{Schlafly} E.~F.,  {Finkbeiner} D.~P.,  2011, \mn@doi [ApJ]
  {10.1088/0004-637X/737/2/103}, \href
  {https://ui.adsabs.harvard.edu/abs/2011ApJ...737..103S} {737, 103}

\bibitem[\protect\citeauthoryear{{Stella}, {White}  \& {Rosner}}{{Stella}
  et~al.}{1986}]{Stella:1986}
{Stella} L.,  {White} N.~E.,   {Rosner} R.,  1986, \mn@doi [ApJ]
  {10.1086/164538}, \href
  {https://ui.adsabs.harvard.edu/abs/1986ApJ...308..669S} {308, 669}

\bibitem[\protect\citeauthoryear{Str{\"{u}}der et~al.,}{Str{\"{u}}der
  et~al.}{2001}]{Struder:2001}
Str{\"{u}}der L.,  et~al., 2001, \mn@doi [A{\&}A] {10.1051/0004-6361:20000066},
  \href {https://ui.adsabs.harvard.edu/abs/2001A&A...365L..18S} {365, L18}

\bibitem[\protect\citeauthoryear{{Subramanian} \& {Subramaniam}}{{Subramanian}
  \& {Subramaniam}}{2013}]{Subramanian:2013}
{Subramanian} S.,  {Subramaniam} A.,  2013, \mn@doi [A{\&}A]
  {10.1051/0004-6361/201219327}, \href
  {https://ui.adsabs.harvard.edu/abs/2013A&A...552A.144S} {552, A144}

\bibitem[\protect\citeauthoryear{{Takata}, {Hu}, {Lin}, {Tam}, {Pal}, {Hui},
  {Kong}  \& {Cheng}}{{Takata} et~al.}{2018}]{Takata:2018}
{Takata} J.,  {Hu} C.~P.,  {Lin} L.~C.~C.,  {Tam} P.~H.~T.,  {Pal} P.~S.,
  {Hui} C.~Y.,  {Kong} A.~K.~H.,   {Cheng} K.~S.,  2018, \mn@doi [ApJ]
  {10.3847/1538-4357/aaa23d}, \href
  {https://ui.adsabs.harvard.edu/abs/2018ApJ...853..106T} {853, 106}

\bibitem[\protect\citeauthoryear{Tiengo, Mereghetti, Turolla, Zane, Rea, Stella
   \& Israel}{Tiengo et~al.}{2005}]{Tiengo:2005}
Tiengo A.,  Mereghetti S.,  Turolla R.,  Zane S.,  Rea N.,  Stella L.,   Israel
  G.~L.,  2005, \mn@doi [A{\&}A] {10.1051/0004-6361:20052633}, \href
  {https://ui.adsabs.harvard.edu/abs/2005A&A...437..997T} {437, 997}

\bibitem[\protect\citeauthoryear{Tiengo, Esposito  \& Mereghetti}{Tiengo
  et~al.}{2008}]{Tiengo:2008}
Tiengo A.,  Esposito P.,   Mereghetti S.,  2008, \mn@doi [ApJ]
  {10.1086/590078}, \href
  {https://ui.adsabs.harvard.edu/abs/2008ApJ...680L.133T} {680, L133}

\bibitem[\protect\citeauthoryear{Tiengo et~al.,}{Tiengo
  et~al.}{2009}]{Tiengo:2009}
Tiengo A.,  et~al., 2009, \mn@doi [MNRAS] {10.1111/J.1745-3933.2009.00728.X},
  \href {https://ui.adsabs.harvard.edu/abs/2009MNRAS.399L..74T} {399, L74}

\bibitem[\protect\citeauthoryear{Turner et~al.,}{Turner
  et~al.}{2001}]{Turner:2001}
Turner M.~J.,  et~al., 2001, \mn@doi [A{\&}A] {10.1051/0004-6361:20000087},
  \href {https://ui.adsabs.harvard.edu/abs/2001A&A...365L..27T} {365, L27}

\bibitem[\protect\citeauthoryear{Turolla, Zane  \& Watts}{Turolla
  et~al.}{2015}]{Turolla:2015}
Turolla R.,  Zane S.,   Watts A.~L.,  2015, \mn@doi [RPPh]
  {10.1088/0034-4885/78/11/116901}, \href
  {https://ui.adsabs.harvard.edu/abs/2015RPPh...78k6901T} {78, 116901}

\bibitem[\protect\citeauthoryear{Vasilopoulos, Haberl, Sturm, Maggi  \&
  Udalski}{Vasilopoulos et~al.}{2014}]{Vasilopoulos:2014}
Vasilopoulos G.,  Haberl F.,  Sturm R.,  Maggi P.,   Udalski A.,  2014, \mn@doi
  [A{\&}A] {10.1051/0004-6361/201423934}, \href
  {https://ui.adsabs.harvard.edu/abs/2014A&A...567A.129V} {567, A129}

\bibitem[\protect\citeauthoryear{Vasilopoulos, Zezas, Antoniou  \&
  Haberl}{Vasilopoulos et~al.}{2017}]{Vasilopoulos:2017}
Vasilopoulos G.,  Zezas A.,  Antoniou V.,   Haberl F.,  2017, \mn@doi [MNRAS]
  {10.1093/mnras/stx1507}, \href
  {https://ui.adsabs.harvard.edu/abs/2017MNRAS.470.4354V} {470, 4354}

\bibitem[\protect\citeauthoryear{Vasilopoulos, Maitra, Haberl, Hatzidimitriou
  \& Petropoulou}{Vasilopoulos et~al.}{2018}]{Vasilopoulos:2018}
Vasilopoulos G.,  Maitra C.,  Haberl F.,  Hatzidimitriou D.,   Petropoulou M.,
  2018, \mn@doi [MNRAS] {10.1093/mnras/stx3139}, \href
  {https://ui.adsabs.harvard.edu/abs/2018MNRAS.475..220V} {475, 220}

\bibitem[\protect\citeauthoryear{Vasisht \& Gotthelf}{Vasisht \&
  Gotthelf}{1997}]{Vasisht:1997}
Vasisht G.,  Gotthelf E.~V.,  1997, \mn@doi [ApJ] {10.1086/310843}, \href
  {https://ui.adsabs.harvard.edu/abs/1997ApJ...486L.129V} {486, L129}

\bibitem[\protect\citeauthoryear{Verner, Ferland, Korista  \& Yakovlev}{Verner
  et~al.}{1996}]{Verner:1996}
Verner D.~A.,  Ferland G.~J.,  Korista K.~T.,   Yakovlev D.~G.,  1996, \mn@doi
  [ApJ] {10.1086/177435}, \href
  {https://ui.adsabs.harvard.edu/abs/1996ApJ...465..487V} {465, 487}

\bibitem[\protect\citeauthoryear{Webb et~al.,}{Webb
  et~al.}{2020}]{Webb4XMM:2020}
Webb N.~A.,  et~al., 2020, \mn@doi [A{\&}A] {10.1051/0004-6361/201937353},
  \href {https://ui.adsabs.harvard.edu/abs/2020A&A...641A.136W} {641, A136}

\bibitem[\protect\citeauthoryear{Wilms, Allen  \& McCray}{Wilms
  et~al.}{2000}]{Wilms:2000}
Wilms J.,  Allen A.,   McCray R.,  2000, \mn@doi [ApJ] {10.1086/317016}, \href
  {https://ui.adsabs.harvard.edu/abs/2000ApJ...542..914W} {542, 914}

\bibitem[\protect\citeauthoryear{Yungelson, Kuranov  \& Postnov}{Yungelson
  et~al.}{2019}]{Yungelson:2019}
Yungelson L.~R.,  Kuranov A.~G.,   Postnov K.~A.,  2019, \mn@doi [MNRAS]
  {10.1093/MNRAS/STZ467}, \href
  {https://ui.adsabs.harvard.edu/abs/2019MNRAS.485..851Y} {485, 851}

\bibitem[\protect\citeauthoryear{Zaritsky, Harris, Thompson  \&
  Grebel}{Zaritsky et~al.}{2004}]{Zaritsky:2004}
Zaritsky D.,  Harris J.,  Thompson I.~B.,   Grebel E.~K.,  2004, \mn@doi [AJ]
  {10.1086/423910}, \href
  {https://ui.adsabs.harvard.edu/abs/2004AJ....128.1606Z} {128, 1606}

\makeatother
\end{thebibliography}
%\bibliography{references_mendeley}

% Alternatively you could enter them by hand, like this:
% This method is tedious and prone to error if you have lots of references
%\begin{thebibliography}{99}
%\bibitem[\protect\citeauthoryear{Author}{2012}]{Author2012}
%Author A.~N., 2013, Journal of Improbable Astronomy, 1, 1
%\bibitem[\protect\citeauthoryear{Others}{2013}]{Others2013}
%Others S., 2012, Journal of Interesting Stuff, 17, 198
%\end{thebibliography}

%%%%%%%%%%%%%%%%%%%%%%%%%%%%%%%%%%%%%%%%%%%%%%%%%%

%%%%%%%%%%%%%%%%% APPENDICES %%%%%%%%%%%%%%%%%%%%%

%\appendix

%\section{Some extra material}

% If you want to present additional material which would interrupt the flow of the main paper,
% it can be placed in an Appendix which appears after the list of references.

%%%%%%%%%%%%%%%%%%%%%%%%%%%%%%%%%%%%%%%%%%%%%%%%%%

% Don't change these lines
\bsp	% typesetting comment
\label{lastpage}
\end{document}